\documentclass[longauth]{aa}

            \usepackage{euclid}
            \usepackage{graphicx}
            \usepackage{placeins}
            
            \usepackage{txfonts}
            \usepackage[]{hyperref}
            \usepackage{xcolor}
            \usepackage{soul}

\begin{document}
               
               \title{\Euclid: Constraining ensemble photometric redshift distributions with stacked spectroscopy\thanks{This paper is published on behalf of the Euclid Consortium.}}

\newcommand{\orcid}[1]{} 
\newcommand{\lastleadauthor}[0]{S.~Camera} 
\author{M.S.~Cagliari$^{1}$\thanks{\email{marina.cagliari@unimi.it}}, B.R.~Granett$^{2}$, L.~Guzzo$^{3,1,2}$, M.~Bolzonella$^{4}$, L.~Pozzetti$^{4}$, I.~Tutusaus$^{5,6}$, S.~Camera$^{7,8,9}$, A.~Amara$^{10}$, N.~Auricchio$^{4}$
% ...
}

\institute{$^{1}$ Dipartimento di Fisica "Aldo Pontremoli", Universit\'a degli Studi di Milano, Via Celoria 16, I-20133 Milano, Italy\\
$^{2}$ INAF-Osservatorio Astronomico di Brera, Via Brera 28, I-20122 Milano, Italy\\
$^{3}$ INFN-Sezione di Milano, Via Celoria 16, I-20133 Milano, Italy\\
$^{4}$ INAF-Osservatorio di Astrofisica e Scienza dello Spazio di Bologna, Via Piero Gobetti 93/3, I-40129 Bologna, Italy\\
$^{5}$ Institut d’Estudis Espacials de Catalunya (IEEC), Carrer Gran Capit\'a 2-4, 08034 Barcelona, Spain\\
$^{6}$ Institute of Space Sciences (ICE, CSIC), Campus UAB, Carrer de Can Magrans, s/n, 08193 Barcelona, Spain\\
$^{7}$ Dipartimento di Fisica, Universit\'a degli Studi di Torino, Via P. Giuria 1, I-10125 Torino, Italy\\
$^{8}$ INAF-Osservatorio Astrofisico di Torino, Via Osservatorio 20, I-10025 Pino Torinese (TO), Italy\\
$^{9}$ INFN-Sezione di Torino, Via P. Giuria 1, I-10125 Torino, Italy\\
$^{10}$ Institute of Cosmology and Gravitation, University of Portsmouth, Portsmouth PO1 3FX, UK\\
$^{11}$ Max Planck Institute for Extraterrestrial Physics, Giessenbachstr. 1, D-85748 Garching, Germany\\
$^{12}$ Universit\"ats-Sternwarte M\"unchen, Fakult\"at f\"ur Physik, Ludwig-Maximilians-Universit\"at M\"unchen, Scheinerstrasse 1, 81679 M\"unchen, Germany\\
$^{13}$ Department of Mathematics and Physics, Roma Tre University, Via della Vasca Navale 84, I-00146 Rome, Italy\\
$^{14}$ INFN-Sezione di Roma Tre, Via della Vasca Navale 84, I-00146, Roma, Italy\\
$^{15}$ INAF-Osservatorio Astronomico di Capodimonte, Via Moiariello 16, I-80131 Napoli, Italy\\
$^{16}$ INAF-IASF Milano, Via Alfonso Corti 12, I-20133 Milano, Italy\\
$^{17}$ Institut de F\'{i}sica d’Altes Energies (IFAE), The Barcelona Institute of Science and Technology, Campus UAB, 08193 Bellaterra (Barcelona), Spain\\
$^{18}$ INAF-Osservatorio Astronomico di Roma, Via Frascati 33, I-00078 Monteporzio Catone, Italy\\
$^{19}$ Department of Physics "E. Pancini", University Federico II, Via Cinthia 6, I-80126, Napoli, Italy\\
$^{20}$ INFN section of Naples, Via Cinthia 6, I-80126, Napoli, Italy\\
$^{21}$ Dipartimento di Fisica e Astronomia “Augusto Righi” - Alma Mater Studiorum Università di Bologna, via Piero Gobetti 93/2, I-40129 Bologna, Italy\\
$^{22}$ INAF-Osservatorio Astrofisico di Arcetri, Largo E. Fermi 5, I-50125, Firenze, Italy\\
$^{23}$ Centre National d'Etudes Spatiales, Toulouse, France\\
$^{24}$ Institut national de physique nucl\'eaire et de physique des particules, 3 rue Michel-Ange, 75794 Paris C\'edex 16, France\\
$^{25}$ Institute for Astronomy, University of Edinburgh, Royal Observatory, Blackford Hill, Edinburgh EH9 3HJ, UK\\
$^{26}$ Jodrell Bank Centre for Astrophysics, Department of Physics and Astronomy, University of Manchester, Oxford Road, Manchester M13 9PL, UK\\
$^{27}$ European Space Agency/ESRIN, Largo Galileo Galilei 1, 00044 Frascati, Roma, Italy\\
$^{28}$ ESAC/ESA, Camino Bajo del Castillo, s/n., Urb. Villafranca del Castillo, 28692 Villanueva de la Ca\~nada, Madrid, Spain\\
$^{29}$ Univ Lyon, Univ Claude Bernard Lyon 1, CNRS/IN2P3, IP2I Lyon, UMR 5822, F-69622, Villeurbanne, France\\
$^{30}$ Mullard Space Science Laboratory, University College London, Holmbury St Mary, Dorking, Surrey RH5 6NT, UK\\
$^{31}$ Department of Astronomy, University of Geneva, ch. d\'Ecogia 16, CH-1290 Versoix, Switzerland\\
$^{32}$ Universit\'e Paris-Saclay, CNRS, Institut d'astrophysique spatiale, 91405, Orsay, France\\
$^{33}$ INAF-Osservatorio Astronomico di Padova, Via dell'Osservatorio 5, I-35122 Padova, Italy\\
$^{34}$ University of Lyon, UCB Lyon 1, CNRS/IN2P3, IUF, IP2I Lyon, France\\
$^{35}$ INAF-Osservatorio Astronomico di Trieste, Via G. B. Tiepolo 11, I-34131 Trieste, Italy\\
$^{36}$ Istituto Nazionale di Astrofisica (INAF) - Osservatorio di Astrofisica e Scienza dello Spazio (OAS), Via Gobetti 93/3, I-40127 Bologna, Italy\\
$^{37}$ Istituto Nazionale di Fisica Nucleare, Sezione di Bologna, Via Irnerio 46, I-40126 Bologna, Italy\\
$^{38}$ Institute of Theoretical Astrophysics, University of Oslo, P.O. Box 1029 Blindern, N-0315 Oslo, Norway\\
$^{39}$ Leiden Observatory, Leiden University, Niels Bohrweg 2, 2333 CA Leiden, The Netherlands\\
$^{40}$ Jet Propulsion Laboratory, California Institute of Technology, 4800 Oak Grove Drive, Pasadena, CA, 91109, USA\\
$^{41}$ Max-Planck-Institut f\"ur Astronomie, K\"onigstuhl 17, D-69117 Heidelberg, Germany\\
$^{42}$ von Hoerner \& Sulger GmbH, Schlo{\ss}Platz 8, D-68723 Schwetzingen, Germany\\
$^{43}$ Institut d'Astrophysique de Paris, 98bis Boulevard Arago, F-75014, Paris, France\\
$^{44}$ Aix-Marseille Univ, CNRS/IN2P3, CPPM, Marseille, France\\
$^{45}$ AIM, CEA, CNRS, Universit\'{e} Paris-Saclay, Universit\'{e} de Paris, F-91191 Gif-sur-Yvette, France\\
$^{46}$ Universit\'e de Gen\`eve, D\'epartement de Physique Th\'eorique and Centre for Astroparticle Physics, 24 quai Ernest-Ansermet, CH-1211 Gen\`eve 4, Switzerland\\
$^{47}$ Department of Physics and Helsinki Institute of Physics, Gustaf H\"allstr\"omin katu 2, 00014 University of Helsinki, Finland\\
$^{48}$ NOVA optical infrared instrumentation group at ASTRON, Oude Hoogeveensedijk 4, 7991PD, Dwingeloo, The Netherlands\\
$^{49}$ Argelander-Institut f\"ur Astronomie, Universit\"at Bonn, Auf dem H\"ugel 71, 53121 Bonn, Germany\\
$^{50}$ Institute for Computational Cosmology, Department of Physics, Durham University, South Road, Durham, DH1 3LE, UK\\
$^{51}$ INFN-Sezione di Bologna, Viale Berti Pichat 6/2, I-40127 Bologna, Italy\\
$^{52}$ California institute of Technology, 1200 E California Blvd, Pasadena, CA 91125, USA\\
$^{53}$ Observatoire de Sauverny, Ecole Polytechnique F\'ed\'erale de Lau- sanne, CH-1290 Versoix, Switzerland\\
$^{54}$ European Space Agency/ESTEC, Keplerlaan 1, 2201 AZ Noordwijk, The Netherlands\\
$^{55}$ Department of Physics and Astronomy, University of Aarhus, Ny Munkegade 120, DK–8000 Aarhus C, Denmark\\
$^{56}$ Centre for Astrophysics, University of Waterloo, Waterloo, Ontario N2L 3G1, Canada\\
$^{57}$ Department of Physics and Astronomy, University of Waterloo, Waterloo, Ontario N2L 3G1, Canada\\
$^{58}$ Perimeter Institute for Theoretical Physics, Waterloo, Ontario N2L 2Y5, Canada\\
$^{59}$ Institute of Space Science, Bucharest, Ro-077125, Romania\\
$^{60}$ Departamento de Astrof\'{i}sica, Universidad de La Laguna, E-38206, La Laguna, Tenerife, Spain\\
$^{61}$ Instituto de Astrof\'isica de Canarias, Calle V\'ia L\'actea s/n, E-38204, San Crist\'obal de La Laguna, Tenerife, Spain\\
$^{62}$ INFN-Sezione di Roma, Piazzale Aldo Moro, 2 - c/o Dipartimento di Fisica, Edificio G. Marconi, I-00185 Roma, Italy\\
$^{63}$ Dipartimento di Fisica e Astronomia “G.Galilei", Universit\'a di Padova, Via Marzolo 8, I-35131 Padova, Italy\\
$^{64}$ INFN-Padova, Via Marzolo 8, I-35131 Padova, Italy\\
$^{65}$ Departamento de F\'isica, Faculdade de Ci\^encias, Universidade de Lisboa, Edif\'icio C8, Campo Grande, PT1749-016 Lisboa, Portugal\\
$^{66}$ Instituto de Astrof\'isica e Ci\^encias do Espa\c{c}o, Faculdade de Ci\^encias, Universidade de Lisboa, Tapada da Ajuda, PT-1349-018 Lisboa, Portugal\\
$^{67}$ Universidad Polit\'ecnica de Cartagena, Departamento de Electr\'onica y Tecnolog\'ia de Computadoras, 30202 Cartagena, Spain\\
$^{68}$ Kapteyn Astronomical Institute, University of Groningen, PO Box 800, 9700 AV Groningen, The Netherlands\\
$^{69}$ Infrared Processing and Analysis Center, California Institute of Technology, Pasadena, CA 91125, USA\\
$^{70}$ INAF-IASF Bologna, Via Piero Gobetti 101, I-40129 Bologna, Italy\\
$^{71}$ Universit\'e de Paris, CNRS, Astroparticule et Cosmologie, F-75013 Paris, France\\
$^{72}$ Space Science Data Center, Italian Space Agency, via del Politecnico snc, 00133 Roma, Italy\\
$^{73}$ Instituto de Astrof\'isica e Ci\^encias do Espa\c{c}o, Universidade do Porto, CAUP, Rua das Estrelas, PT4150-762 Porto, Portugal\\
}

               \date{}
            
              \abstract
               {The ESA \Euclid mission will produce photometric galaxy samples over 15\,000 square degrees of the sky that will be rich for clustering and weak lensing statistics. The accuracy of the cosmological constraints derived from these measurements will depend on the knowledge of the underlying redshift distributions based on photometric redshift calibrations.}
               {A new approach is proposed to use the stacked spectra from \Euclid slitless spectroscopy to augment broad-band photometric information to constrain the redshift distribution with spectral energy distribution fitting. The high spectral resolution available in the stacked spectra complements the photometry and helps to break the colour-redshift degeneracy and constrain the redshift distribution of galaxy samples.}
               {We modelled the stacked spectra as a linear mixture of spectral templates. The mixture may be inverted to infer the underlying redshift distribution using constrained regression algorithms. We demonstrate the method on simulated Vera C. Rubin Observatory and \Euclid mock survey data sets based on the Euclid Flagship mock galaxy catalogue.  We assess the accuracy of the reconstruction by considering the inference of the baryon acoustic scale from angular two-point correlation function measurements.}
               {We selected mock photometric galaxy samples at redshift $z>1$ using the self-organising map algorithm. Considering the idealised case without dust attenuation, we find that the redshift distributions of these samples can be recovered with 0.5\% accuracy on the baryon acoustic scale. The estimates are not significantly degraded by the spectroscopic measurement noise due to the large sample size.  However, the error degrades to 2\% when the dust attenuation model is left free. We find that the colour degeneracies introduced by attenuation limit the accuracy considering the wavelength coverage of \Euclid near-infrared spectroscopy. }
               {}
            
                \titlerunning{\Euclid: Redshift distribution with stacked spectroscopy}
                \authorrunning{M.S. Cagliari et al.}
               \maketitle
            %

            \section{Introduction} \label{introduction}
            
            The next generation of photometric surveys will produce unprecedented galaxy statistics that will fuel large-scale structure studies \citep{lsst, euclid, jpas, des}. Compared with their spectroscopic counterparts \citep{LeFevre2005,GAMA,VIPERS,DESI}, photometric surveys go deeper and faster; however, the surveying efficiency comes at the cost of spectral resolution. Imaging surveys are limited to photometric measurements such as broad-band colours to infer the redshifts of galaxies \citep{Connolly1995,Bolzonella2000,Benitez2000}. The minimum error in a photometric redshift estimate with optical and near-infrared broad-band photometry is $\sigma_z/(1+z) \sim 0.02$ due to fundamental degeneracies in the colour-redshift parameter space \citep{2019NatAs...3..212S}. Nevertheless, with the promise of large sample sizes, this precision is often acceptable for large-scale structure studies based on galaxy clustering and weak lensing analyses. The redshift of individual galaxies is not required for these analyses, but instead precise knowledge of the redshift distribution of the sample is needed to properly interpret the statistics. A systematic error in the redshift distribution estimate propagates directly to biases in the results \citep{Newman2015}.
            
            The ensemble redshift distribution of a photometrically selected galaxy
            sample can be measured directly by targeting a representative sub-sample with spectroscopy. Currently the complete calibration of the colour-redshift relation (C3R2) campaign is underway using 8m class telescopes to construct a calibration dataset for \Euclid\footnote{\url{http://www.euclid-ec.org}} (\citealp{c3r2}; \citealp{Masters2020}; \citealp{Stanford2012}). It is challenging to build fully representative spectroscopic samples particularly at the faint end at both low and high redshift.  For past surveys it was necessary to include corrections for incompleteness in the spectroscopic measurements \citep{Lima2008,Hartley2020} and these corrections depend on a complex set of parameters related to observing conditions and intrinsic galaxy properties \citep{Scodeggio2018}. 
            An alternative solution, the clustering redshift estimator, uses the signal encoded in the spatial correlation between a photometric sample and reference spectroscopic samples to infer the redshift distribution of the photometric sample \citep{Schneider2006,Newman2008,Schmidt2013,Scottez2016}. This approach is expected to be  successful when applied to the \Euclid data set. Each method to calibrate photometric redshift distributions comes with its own assumptions and sources of systematic errors; therefore, it is worthwhile to develop complementary methods that can provide robust cross-checks. We focus here on an approach that will be enabled by the rich data set provided by \Euclid's slitless spectroscopy.
            
            Slitless spectroscopy provides a unique tool since a measurement of spectral flux can be extracted for every source detected in imaging \citep[e.g.][]{Zwicky1941,3DHST}. The ESA \Euclid mission will be the first modern all-sky survey programme to employ a slitless spectrograph (\citealp{euclid}; SPHEREx, \citeauthor{spherex} \citeyear{spherex}, and the NASA \textit{Nancy Grace Roman}, \citeauthor{roman} \citeyear{roman}, missions will follow). By design the \Euclid spectroscopy will detect and measure the redshifts for the brightest emission line galaxies primarily using the H$\alpha$ line in the redshift range $0.9<z<1.8$.
            The majority of photometrically detected sources will be fainter and give only a very low signal-to-noise ratio spectrum precluding a direct redshift measurement. However, by stacking the spectra we can extract physical information from the ensemble and augment photometric studies.

            The ensemble photometric redshift method proposed by \citet{2019arXiv190301571P}
            aims to constrain the redshift distribution of a photometrically selected galaxy sample by using the stacked spectrum built from the average of many low signal-to-noise ratio spectra. Since broad-band photometric measurements have coarse wavelength resolution,  galaxies with different spectral types at different redshifts can have degenerate colours. These degeneracies lead to catastrophic photometric redshift errors which are characterised by multiple peaks and long tails in the redshift distribution. Adding information from stacked spectroscopy can break these degeneracies since spectral features leave their signature in the stack. The spectra cannot be used to infer the redshift of individual sources due to the low signal-to-noise ratio of the measurements, but the ensemble can be used to infer the redshift distribution.  The stacked spectrum will be a mixture of galaxy spectral types at different redshifts and with template fitting a unique decomposition may be found to recover the redshift distribution.

            In this work we implement and test the approach on a mock galaxy survey considering ground based photometry from the Vera C. Rubin Observatory and near-infrared photometry and slitless spectroscopy from \Euclid. We select galaxy samples based on the photometry and infer the redshift distributions using the combination of stacked spectroscopy and stacked photometry.
            The \Euclid near-infrared spectrograph (NISP) has a wavelength range 1.25--1.85 \micron; therefore, it will measure the rest-frame spectral energy distribution (SED) at $\lambda<9000\, \AA$ for galaxies at $z>1$. Thus the spectroscopy can augment the near-infrared photometry by adding continuum shape information. The 4000\,$\AA$ break which is a key feature for redshift estimation will be accessible for galaxies at $z>2$. 
            The redshift distributions inferred from broad-band photometry alone are generally not accurate because of the dependence on the template priors \citep{Benitez2000}.  However, the joint fit of spectroscopy and photometry together proves to be a powerful tool for extracting redshift distributions: broad-band photometry  provides a broad wavelength coverage and spectroscopy gives higher spectral resolution that can break colour degeneracies. We  demonstrate this in the case of \Euclid in Appendix \ref{sp_vs_p}.  We quantify the accuracy of the constraints by considering the inference of the baryon acoustic oscillation (BAO) scale from angular two-point correlation function measurements. 
            The BAO scale is not the only feature that will be measured in photometric galaxy clustering analyses; the full shape of the galaxy power spectrum encodes relevant information for cosmological studies. However, we can consider that the uncertainty on the BAO scale provides a lower limit on the information contained in the power spectrum and thus is a useful metric for quantifying systematic errors.
            This metric is also applicable to weak lensing studies that require determinations of the mean redshift of tomographic bins.
            

            This paper is organised as follows.
            In Sect. \ref{ensamble_photo-zs}, we present the ensemble photometric redshift method and describe how we reduce the formal problem of finding the redshift distribution of a group of galaxies with similar colours to a linear problem.
            Then, in Sect. \ref{data}, we describe the construction of the mock catalogues used to test the method (Sect. \ref{mock_data}).
            In this section we also discuss the SED templates used to fit the redshift distributions, how the galaxies are partitioned into colour groups and the quantitative benchmark for the redshift distribution estimates based on the measurement of the BAO scale.
            Finally, in Sect. \ref{results}, we present the results of the analyses of both ideal noiseless spectroscopy data and realistic cases with noise.
            In Sect. \ref{conclusions} we summarise our results and discuss the applications and possible improvements that may be made.

            \section{The ensemble photometric redshift method} \label{ensamble_photo-zs}
            
            \begin{figure*}[t]
                \centering
                \includegraphics[width=0.5\textwidth]{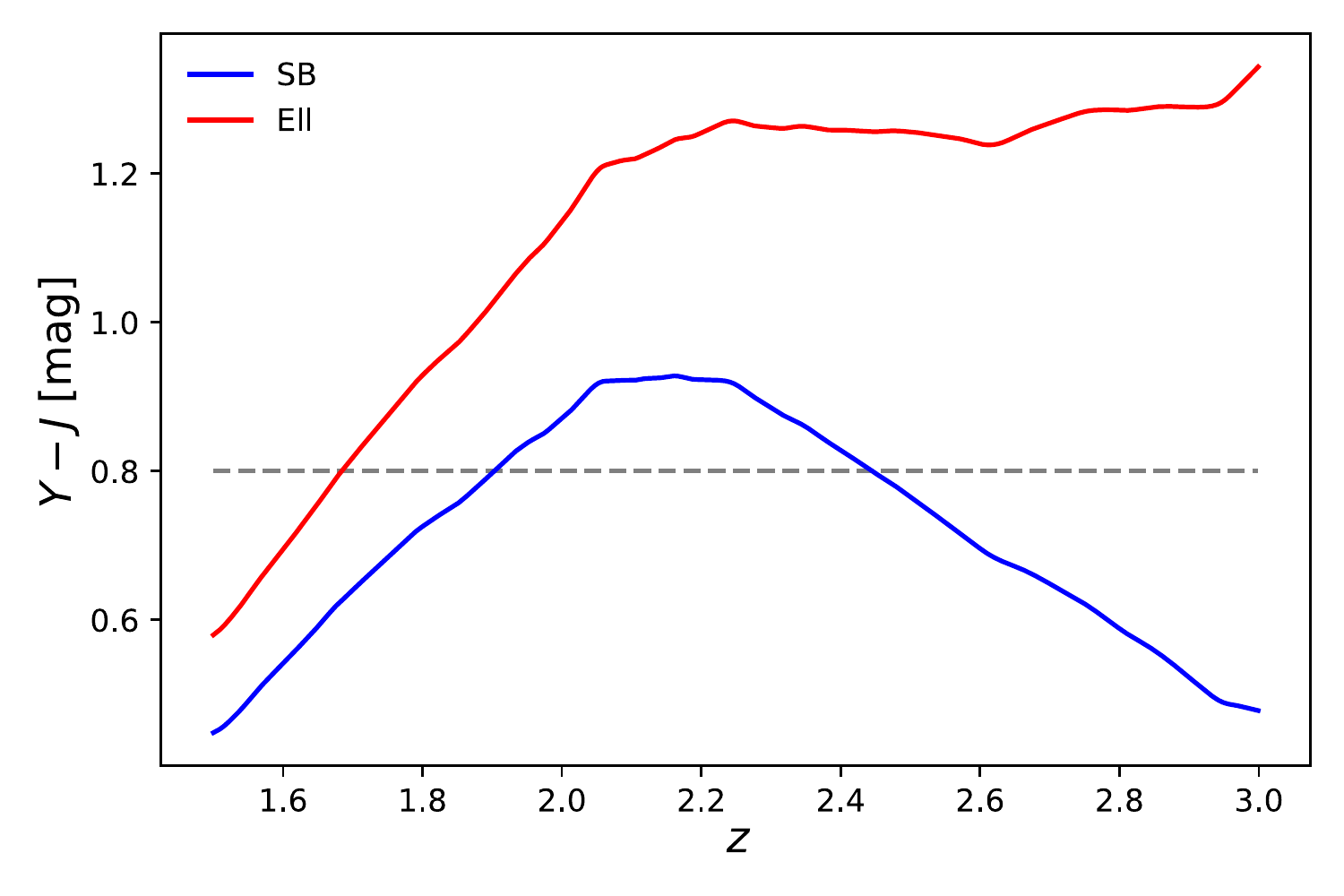}\includegraphics[width=0.5\textwidth]{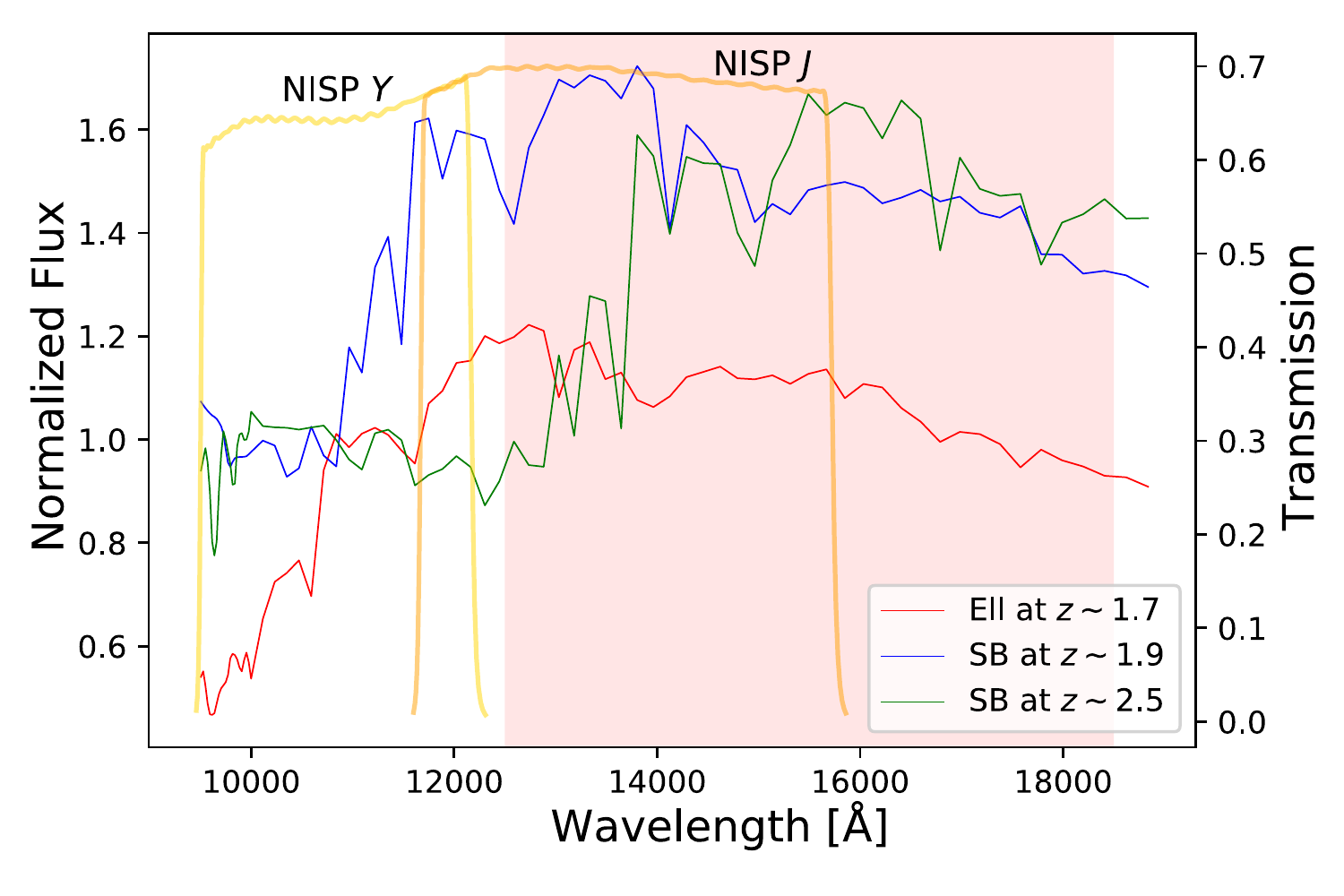}
                \caption{Example of how stacked spectroscopy breaks the colour-redshift degeneracy. \emph{Left}: To illustrate the method we show a colour degeneracy in $Y-J$ as a function of redshift for starburst (SB) and elliptical (Ell) galaxy templates from \citet{2009ApJ...690.1236I}. The dashed horizontal line indicates $Y-J=0.8$ and shows three redshift solutions consistent with the colour: elliptical galaxies at $z \sim 1.7$, and starburst galaxies at $z \sim 1.9$ and $z \sim 2.5$. 
                \emph{Right}: We see that each solution gives a unique spectral shape in the near infrared range probed by the \Euclid NISP instrument (red shaded area). The red line shows the elliptical template at $z \sim 1.7$ and the blue and green lines show the starburst template at $z \sim 1.9$ and $2.5$. The spectra are normalised at the effective wavelength of the $Y$ NISP filter. The \Euclid NISP $Y$ and $J$ filter transmission are overplotted. The spectral resolution of the plotted templates is lower than the \Euclid NISP spectrograph one. The stacked spectroscopy at fixed colour is built from the linear combination of these templates and encodes enough information to recover the relative contributions of spectral type at each redshift.
                }
                \label{fig:breaking_deg}
            \end{figure*}

            \subsection{Method}\label{method}
            

            The distribution of galaxies in colour-redshift space can be constrained by adding information from stacked spectroscopy. This is illustrated in Fig. \ref{fig:breaking_deg}. The left panel shows a three-fold degeneracy in colour at $Y-J=0.8$ for starburst and elliptical galaxy spectral types.
            This  colour can correspond to a population of elliptical galaxies at $z \sim 1.7$, starburst galaxies either at $z \sim 1.9$ or $z \sim 2.5$, or to different mixes of these populations. In the right panel we show that these galaxy populations have unique spectral shapes, and so, the stacked spectrum encodes information about the distribution of redshifts and the mixture of galaxy types. We now consider the general case with many photometric colour measurements and describe how the information in the stacked spectroscopy can be extracted.

            The normalised stacked spectrum (hereafter stacked spectrum) of a sample of galaxies with similar colours (hereafter a colour group) is defined as the weighted mean of the individual spectral flux measurements,
            \begin{equation}
                f_{\rm stack}^{\rm obs}(\lambda) = \frac{1}{N_{\rm gal}} \sum_{i=1}^{N_{\rm gal}} \frac{1}{\bar{f}_i}f_i(\lambda) \,,
            \label{stacked_spectrum}
            \end{equation}
            where $i$ indexes the galaxies, $f(\lambda)$ is the measured galaxy flux as a function of wavelength and $\bar{f}$ is the integrated flux for normalisation, therefore the normalised stacked spectrum is expressed in units ${\rm Hz}^{-1}$.
            The calculation of the integrated flux will be discussed in Sect. \ref{normalization}.
            In the analysis we generalise $f(\lambda)$ to also include broad-band photometric measurements (see Sect. \ref{combining}).
            The observed flux spectrum can be written in terms of the rest-frame SED of the galaxy,
            \begin{equation}
            f (\lambda) = a \  T (\lambda\ |\ z) \,,
            \end{equation}
            thus, as the product of a flux normalisation, $a$, and a rest-frame template transformed to redshift $z$, $T (\lambda\ |\ z)$.
            
            Galaxy SEDs can be modelled by a finite set of parameters \citep[e.g.][]{Marchetti2013}.
            Therefore, the expression for the stacked spectrum can  be rewritten as a sum over discrete SEDs indexed by $\alpha$ and weighted by their frequency as a function of redshift, $p_\alpha(z)$,
            \begin{equation}
                f_{\rm stack}^{\rm model}(\lambda) =  \sum_{\alpha = 1}^{N_{\rm SED}} \int \diff z \  p_\alpha(z) \,  T_\alpha (\lambda\ |\ z)\,.
            \end{equation}
            The template normalisation required to equate $f_{\rm stack}^{\rm obs}(\lambda) =
            f_{\rm stack}^{\rm model}(\lambda)$ will be discussed in Sect. \ref{normalization}.
            
            In order to carry out the numerical analysis we discretise the expression over
            a regular grid of redshift,
            \begin{equation}
                f_{\rm stack}^{\rm model}(\lambda) = \sum_{\alpha = 1}^{N_{\rm SED}} \sum_{z = z_{\rm min}}^{z_{\rm max}} \, p_{\alpha, z} T_{\alpha, z} (\lambda)\,,
            \label{stacked_spectrum_sums}
            \end{equation}
            where $N_{\rm SED}$ is the number of templates, and $z_{\rm min}$ and $z_{\rm max}$ are the minimum and maximum redshift over which to evaluate the redshift distribution.
            The overall redshift distribution is therefore given by the summation over templates
            \begin{equation}
                p_z \propto \sum_{\alpha=1}^{N_{\rm SED}} p_{\alpha, z}\,.
            \label{redshift_distribution}
            \end{equation}
            In principle the redshift distribution can be found by substituting the observed flux $f_{\rm stack}^{\rm obs}$ for $f_{\rm stack}^{\rm model}$ in Eq. (\ref{stacked_spectrum_sums}) and solving for the coefficients $p_{\alpha, z}$.
            
            \subsection{Machine implementation} \label{machine_implementation}
            
            Equation (\ref{stacked_spectrum_sums}) describes a linear set of equations that can be
            written in matrix notation as
            \begin{equation}
                \mathbf{f} = \mathbf{T} \mathbf{p}\,,
            \label{linear_problem}
            \end{equation}
            where $\mathbf{f}$ is the spectral flux data vector with $N_\lambda$ elements.
            The matrix $\mathbf{T}$ is constructed from $N_{\rm SED}$ SED templates each shifted
            to $N_z$ redshifts; therefore $\mathbf{T}$ has dimension $(N_{\rm SED} \, N_z)\times N_\lambda$.
            The redshift distribution of each template is encoded in the vector $\mathbf{p}$ which has length $N_{\rm SED} \, N_z$.
            Since the product of templates and redshifts $N_{\rm SED} \, N_z$ is much greater than the number of spectral elements $N_\lambda$, the system is under-constrained and does not have a unique solution.
            
            We can make progress in solving Eq. (\ref{linear_problem}) by using a linear regression algorithm that employs regularisation terms to identify the most interesting solutions.
            We add two pieces of information: first, we are not interested in unphysical solutions with negative $p_{\alpha,z}$, and second, galaxy spectra are well fitted by a small number of SED templates.
            These two considerations led us to impose a non-negativity constraint and to use a shrinkage estimator\footnote{In statistics, shrinkage is a process that aims to reduce overfitting and the effect of sampling variation. It can be implemented with the addition of penalties to the cost function of interest.} to find the minimum set of templates that can fit the stacked spectra.
            
            We tested three linear regression methods with non-negativity constraints:
            %
            first, the non-negative least squares method \citep[NNLS,][]{nnls}, second, the least absolute shrinkage and selection operator \citep[LASSO,][]{10.2307/2346178}, and last, the elastic net regularisation \citep[ElasticNet,][]{doi:10.1111/j.1467-9868.2005.00503.x}.
            All three methods minimise a cost function of the form
            \begin{equation}
                \min_{p \ge 0} \left\{ \frac{1}{N} \sum_{i} \left( f_i - \sum_j T_{ji} p_j \right)^2 + Q(p_i) \right\}\,,
            \label{least_squares}
            \end{equation}
            but adopt different penalty functions $Q$.
            The penalty function acts to reward solutions that use fewer templates.
            LASSO adds the $l_1$ penalty of the form $Q(p,\alpha) = \alpha |p| $ and ElasticNet uses $Q(p,\alpha,\beta) = \alpha \left[\beta |p| + 0.5 (1-\beta) p^2 \right]$.
            The variables $\alpha$ and $\beta$ are free parameters that must be chosen in the analysis (see Sect. \ref{parameter_selection}).
            NNLS is a variant of the standard least squares solver and does not introduce a penalty term.
            In this work we used the implementation of the NNLS algorithm in the Python SciPy library {\tt optimize.nnls}  \citep{2020SciPy-NMeth}.
            For LASSO and ElasticNet we used the implementations found in the Scikit-learn library \citep{JMLR:v12:pedregosa11a}.

            The attractiveness of the ensemble photometric redshifts method as it has been presented here comes from its ability to infer the underlying distribution using only a chosen template set and no additional information. However, adding physical priors, for example the galaxy luminosity function or galaxy type-redshift distributions,  may improve the method performance. Considering how the problem was reduced to a set of linear equations (Eq. \ref{linear_problem}), to take into account physical priors is not a trivial task. Possibly, the most straightforward way to do so is to rewrite the problem in terms of likelihood maximisation in a Bayesian framework. The likelihood could be sampled in the parameter space via a Markov chain Monte Carlo (MCMC) algorithm. This approach could have a high computational cost since the parameter space is very large.
            
            \subsection{Normalisation}\label{normalization}
            
            The normalisation of the spectra is important in the stacking process (Eq. \ref{stacked_spectrum}) to standardise the contribution from the faintest and brightest sources.
            We chose to normalise the galaxy spectroscopy by the integrated flux.
            However, since the measured spectra are very noisy, the integration cannot be carried out on the spectra themselves.
            Instead, we used the broad-band photometry to set the normalisation.
            The photometry is typically deeper than the spectroscopy and so gives a robust normalisation.
            
            In this analysis we used the near-infrared photometry in the $Y$, $J$ and $H$ bandpasses that will be measured by the \Euclid NISP instrument.
            That is, the integrated flux used to weight the measured spectra in Eq. (\ref{stacked_spectrum}) is given by
            \begin{equation}
                \bar{f} = f_{\rm Y} + f_{\rm J} + f_{\rm H}\,,
            \label{spectrum_normalization}
            \end{equation}
            where $f_{\rm Y}$, $f_{\rm J}$ and $f_{\rm H}$ represent the measured photometric flux in the $Y$, $J$ and $H$ bandpasses.
            
            The SED templates, $T_{\alpha, z}(\lambda)$, were normalised in the same way by computing the integrated flux in the three NISP bandpasses and summing them.
            The flux integrated over a bandpass response function $R(\lambda)$ is
            \begin{equation}
                f_R = \frac{\int  R(\lambda)\, f_\lambda(\lambda)\, \frac{\lambda}{hc}\, \diff \lambda}{\int  R(\lambda)\, \frac{\diff \lambda}{h \lambda} } \, ,
            \label{integrated_flux}
            \end{equation}
            where $f_\lambda(\lambda)$ is the spectral flux in units ${\rm erg\,cm^{-2}\,s^{-1}\,\AA^{-1}}$ and $hc$ is the product of Planck's constant and the speed of light.
            
            \subsection{Combining photometry and spectroscopy} \label{combining}
            
            The extension of the observed stacked spectrum $\mathbf{f}$ with photometry is straightforward.
            We generalised the spectroscopic wavelength $\lambda$ in Eq. (\ref{stacked_spectrum}) so that it also referred to photometric bands.
            The first part of the data vector $\mathbf{f}$ contains the actual stacked spectrum, while its last $N_{\rm b}$ elements, where $N_{\rm b}$ is the number of photometric filters, is the observed stacked photometry in each filter.
            The photometric data were stacked following Eq. (\ref{stacked_spectrum}) in the same way as the spectra were and have the same normalisation as well (Eq. \ref{spectrum_normalization}). Therefore, the extended stacked spectrum $\mathbf{f}$ is a vector with $N_{\lambda} + N_{\rm b}$ components.
            
            In order to extend the template matrix with photometry  we computed the photometric fluxes in the bandpasses of interest with Eq. (\ref{integrated_flux}) for each one of the $N_{\rm SED} \, N_z$ templates in the matrix.
            These fluxes were normalised in the same way as the SED templates.
            The dimension of the template matrix $\mathbf{T}$ becomes $(N_{\rm SED} \, N_z) \times (N_{\lambda} + N_{\rm b})$ and its columns have the same order as the elements of the extended stacked spectrum.
            
            Lastly, in this work we do not weight the data with their observational errors.
            This choice was dictated by the great number of galaxies in the colour groups.
            There are so many galaxies in a colour group that the noise in the spectra becomes negligible. This is seen in Fig. \ref{fig:stack} right panel which illustrates the spectral stack with $2\times10^6$ galaxies.
            However, an inverse-error weighting may be applied to improve the performance in the analysis of less populous colour groups.
            The weights may be defined using the variance of the stacked spectrum,
            \begin{equation}
                \sigma_{\rm stack}^2 (\lambda) = \frac{1}{N_{\rm gal}^2} \sum_{i = 1}^{N_{\rm gal}} \frac{1}{\bar{f_i}^2}\, \sigma_i^2 (\lambda)\,,
            \label{stacked_error}
            \end{equation}
            where $\sigma(\lambda)$ is the observed galaxy flux error as a function of wavelength.
            The stacked standard deviation is the square root of the stacked variance and its inverse can be used to weight the stacked spectrum and the columns of the template matrix.
            If the data are weighted following this recipe, the computation of $\mathbf{p}$ remains a linear problem with the form of Eq. (\ref{linear_problem}) with the only difference being that we substitute  the stacked spectrum and the template matrix with their weighted counterparts.
            
            \section{Application to mock \Euclid data}\label{data}
            
            \subsection{Survey simulation}\label{mock_data}
            
            We synthesised mock spectroscopic and photometric observations representative of the \Euclid survey to validate the ensemble photometric redshifts method.
            We based the simulations on the Euclid Flagship mock galaxy catalogue v1.8.4\footnote{\url{https://sci.esa.int/web/euclid/-/59348-euclid-flagship-mock-galaxy-catalogue}}.
            The Euclid Flagship simulation is a dark matter N-body simulation with a box size of $3780\,h^{-1}\, \rm Mpc$ and particle mass of $2.4 \times 10^9 \, \Msolar$ \citep{2017ComAC...4....2P}.
             
            The cosmic web of dark matter halos in the Flagship simulation was populated with galaxies using an extended halo occupation distribution model \citep{Carretero2015,2015MNRAS.453.1513C} by the SciPIC collaboration \citep{inproceedings,TALLADA2020100391} and a full-sky light cone was produced spanning the redshift range from $0$ to $2.3$.
            Galaxy properties, including the SEDs and broad-band magnitudes were assigned to match the luminosity function and galaxy clustering measurements at $z=0.1$ and extrapolated to higher redshift.

            In this work we used the Flagship catalogue v1.8.4 covering one octant of the sky (\num{5157}$\,\mathrm{deg^2}$) in the redshift range $0 < z < 2.3$.
            We used the SED of each galaxy to simulate the \Euclid grism spectroscopy in the near infrared as well as the \Euclid broad-band photometry $Y$, $J$ and $H$ bands and the six bands from the Vera C. Rubin Observatory: $u$, $g$, $r$, $i$, $z$ and $y$\footnote{The \Euclid and Vera C. Rubin Observatory filter transmission functions were obtained from the Euclid data model version 1.8.}.
            The flux from spectral lines was not simulated in the SEDs or broad-band photometry. This choice simplified the SED fitting procedure but is an idealisation that should be addressed in a future analysis. However, the addition of emission line flux on the photometry is minor compared with the effect of internal attenuation which we describe next.
            
            The attenuated mock SEDs are based on the COSMOS template set \citep{2009ApJ...690.1236I} with a variation in the internal galaxy attenuation curves with the addition of the $2175 \, \AA$ bump \citep{1984A&A...132..389P,2000ApJ...533..682C}. Each mock SED has an associated attenuation model that comes from matching against the COSMOS catalogue.
            We used three mock catalogue versions in the analysis with different attenuation models:
            \begin{enumerate}
              \item `non-attenuated' - galactic attenuation was not applied to the SEDs;
              \item `fixed' - a fixed attenuation model was applied to all galaxy SEDs (see below);
              \item `real' - the value of $E(B-V)$ for each galaxy in the Flagship catalogue was used to apply attenuation to the SED.
            \end{enumerate}
            In each case the broad-band photometry was computed in a consistent way from the SED with Eq. (\ref{integrated_flux}).
            
            We applied attenuation to the SED in the following way that is consistent with the construction of the Flagship mock galaxy catalogue. The attenuated SED was computed as the product of the non-attenuated SED and an attenuation factor,
            \begin{equation}
                F_{\rm att}(\lambda) = \left( \frac{f_{\rm att}(\lambda)}{f_0(\lambda)} \right)^{\frac{E(B-V)}{0.2}},
            \label{extinction_factor}
            \end{equation}
            where $f_{\rm att}(\lambda)$ is one out of the four attenuation curves from \citet{1984A&A...132..389P} and \citet{2000ApJ...533..682C}, $f_0(\lambda)$ is a constant function per unit frequency and $E(B-V)$ is the colour excess.
            We note that the parametrisation in Eq. (\ref{extinction_factor}) is peculiar to the construction of the Flagship catalogue. 
            To build the fixed attenuation catalogue all of the galaxy SEDs were multiplied by the same attenuation factor computed with the attenuation curve from \citet{1984A&A...132..389P} and $E(B-V) = 0.2$.
            In the case of the real attenuation catalogue, the attenuation curve and the value of $E(B-V)$ specified for each galaxy in the Flagship catalogue were used.
    
            We simulated the measurement uncertainty of the mock spectroscopy and photometry using a simple photometric model.
            The signal-to-noise ratio ($\mathrm{S/N}$) was defined as
            \begin{equation}
                \mathrm{S/N} = \frac{f}{\sigma}\,,
            \label{snr}
            \end{equation}
            where $f$ is the band flux and $\sigma$ its measurement uncertainty.
            Then the variance of a given flux can be computed as
            \begin{equation}
                \sigma_f^2 = \frac{f_{\rm lim}}{\mathrm{S/N}_{\rm lim}^2} \,f\,,
            \label{flux_variance}
            \end{equation}
            where $f_{\rm lim}$ is the flux corresponding to the instrumental depth in the chosen band, $\mathrm{S/N}_{\rm lim}$ is the signal-to-noise ratio at which the depth is expressed and $f$ is the true galaxy flux.
            We adopted the $10 \, \sigma$ depth values (J.C. Cuillandre, private communication) listed in AB magnitudes in Table \ref{tab:fluxlim}.
            We generated the observed photometric flux by drawing a value from a Gaussian distribution centred on the real photometric flux and with standard deviation $\sigma_f$.
            Moreover, in order to simulate the measured galaxy sample we applied an $H$-band magnitude selection $H<24$ and a signal-to-noise ratio threshold of $\mathrm{S/N_H} > 5$ for the redshift distribution analysis.
            The total number of galaxies in each catalogue was about $10^9$ after the signal-to-noise ratio selection.
            \begin{table}
                \caption{$10\, \sigma$ depths.}
                \centering
                \begin{tabular}{c c}
                \hline
                \\ [-1em]
                     Filter & $10\, \sigma$ depth \\ \hline
                     \\ [-1em]
                     $u$ &  $24.2$ \\
                     $g$ & $24.5$ \\
                     $r$ & $23.9$ \\
                     $i$ & $23.6$ \\
                     $z$ & $23.4$ \\
                     $y$ & $23.2$ \\
                     $Y$ & $23.0$ \\
                     $J$ & $23.0$ \\
                     $H$ & $23.0$ \\ \hline
                \end{tabular} 
                \vspace{1ex}

            {\raggedright \small \textbf{Notes.} The $10\, \sigma$ depth values in AB magnitude adopted for each filter. \par}
            \label{tab:fluxlim}
            \end{table}
            
            The \Euclid NISP spectrograph is sensitive over the wavelength range $1.25 < \lambda < 1.85 \, \micron$.
            The pixel dispersion is $\Delta_\lambda=13.4 \, \AA \,{\rm pixel}^{-1}$ such that the spectral data  vector has $N_\lambda=488$ elements.
            We modelled the measurement uncertainty  with instrumental and astrophysical noise sources.
            The variance on the detector in electron count units per pixel is
            \begin{equation}
                \sigma^2_{\rm pixel} = N_{\rm exp} \left[ t_{\rm exp}  \left(n_{\rm dark} + n_{\rm sky}\right) + \sigma^2_{\rm read} \right]\,,
            \end{equation}
            where $N_{\rm exp}$ is the number of exposures, $t_{\rm exp}$ is the exposure time.
            The detector noise has contributions from the dark current $n_{\rm dark}$ and the read noise $\sigma^2_{\rm read}$.
            The astrophysical background $n_{\rm sky}$ includes contributions from zodiacal emission and scattered light.
            The noise per pixel was propagated to the flux-calibrated one-dimensional spectrum $\sigma_\lambda(\lambda)$  in units ${\rm erg\,cm^{-2}\,s^{-1}\,\AA^{-1}}$ with
            \begin{equation}
                \sigma_\lambda(\lambda) =  \frac{h c \, \lambda \, w}{A\, \Delta_\lambda \,q_e(\lambda) \,T(\lambda)} \,\sigma_{\rm pixel}\,.
            \end{equation}
            Here, $A$ is the collecting area of the telescope, $\Delta_\lambda$ is the spectral dispersion in $\rm \AA \, pixel^{-1}$, $w$ is the extraction window in pixels, $q_e$ is the detector quantum efficiency and $T$ is the transmission function.
            The measurement uncertainty was assigned to the model flux spectra by computing the spectral variance $\sigma_{\lambda}^2(\lambda)$.
            The noisy realisations of the spectra were generated by drawing values from a Gaussian distribution with the given variance and adding them to the model flux spectra. 
            
            \begin{figure*}
                \centering
                \includegraphics[width=0.5\textwidth]{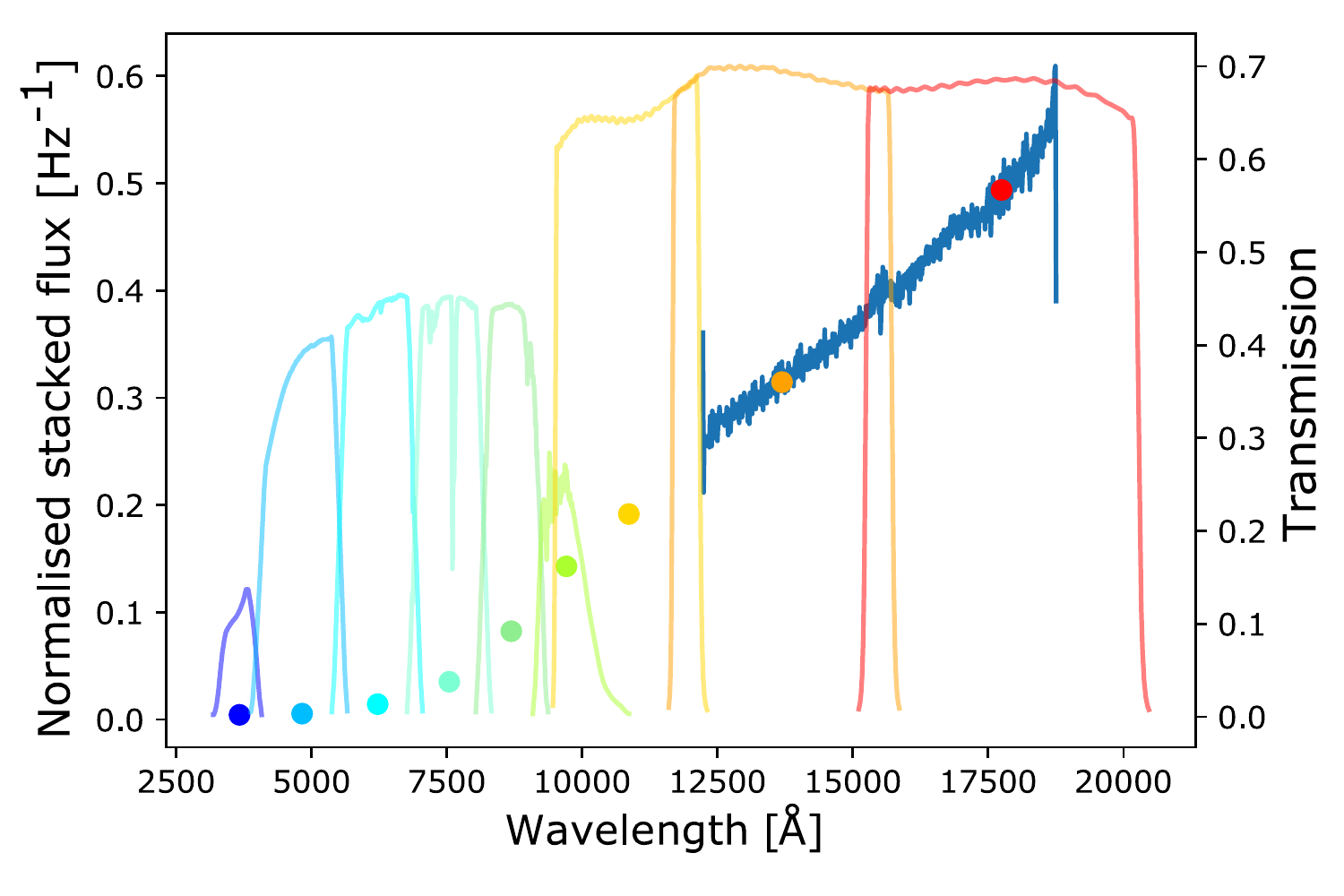}\includegraphics[width=0.5\textwidth]{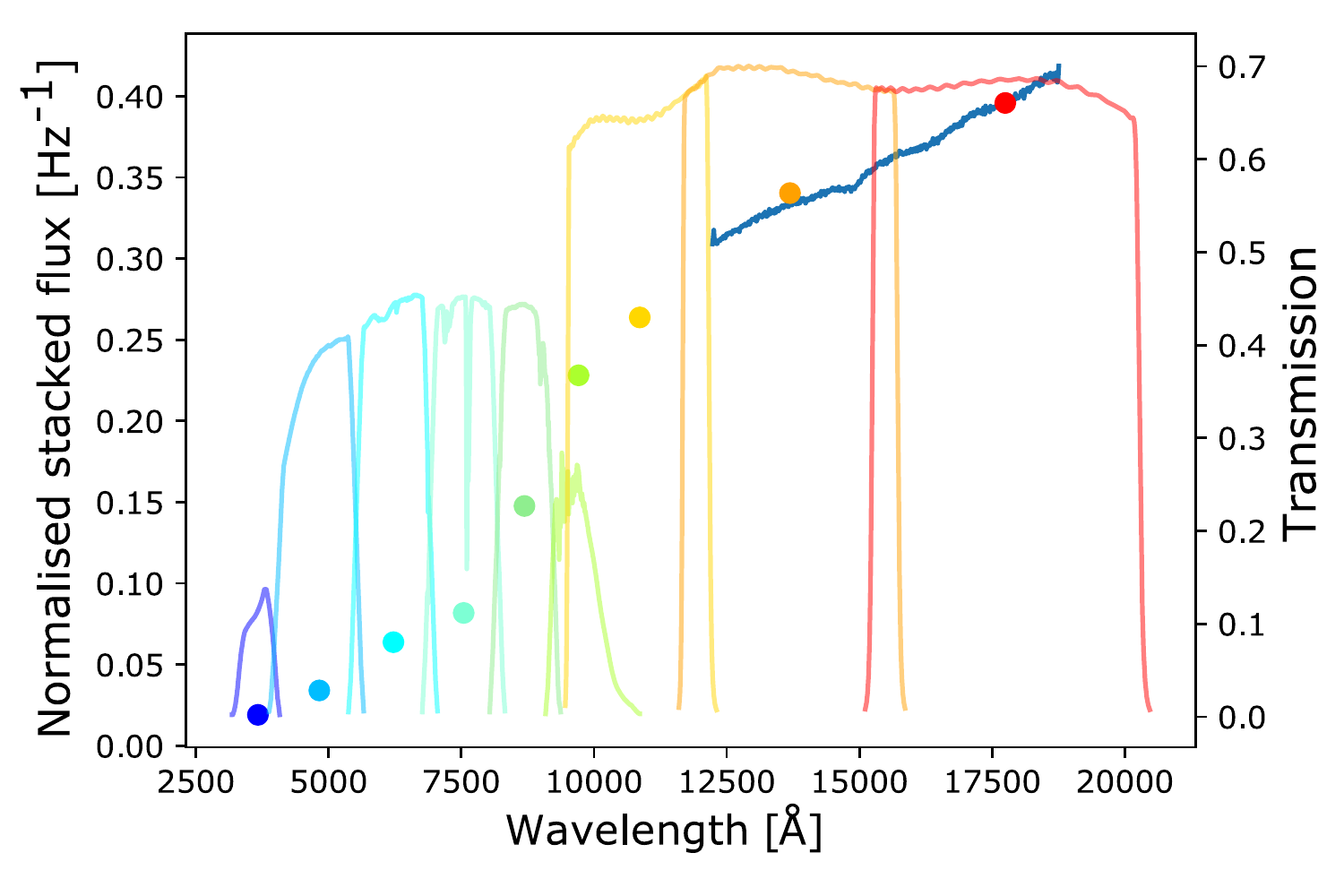}
                \caption{
                Stacked spectro-photometry for two groups of galaxies that have been photometrically selected (see Sect. \ref{sec:SOM}). The photometric data includes \Euclid $Y J H$ and the Vera C. Rubin Observatory $ugrizy$ bands and the spectrocscopic data is from the \Euclid NISP instrument with simulated noise. The photometric bandpasses used in the analysis are over-plotted. In both plots the photometric uncertainty bars are smaller than the markers. \emph{Left}: an example stack for a group of galaxies with mean redshift $z=1.10$. The stack includes  $2 \times 10^3$ galaxies. \emph{Right}: stacked flux for a group at mean redshift $z=1.44$ with $ 2\times 10^6$ galaxies.
                }
                \label{fig:stack}
            \end{figure*}
            In Fig. \ref{fig:stack} we show an example stacked spectrum traced by $ugrizyYJH$ photometry and NISP spectroscopic measurements. The left and right panels show stacks built from $2\times10^3$ and $2\times10^6$ sources, respectively. The measurement uncertainty on the photometric points is not visible in both cases while the uncertainty on the spectroscopy is evident. The spectroscopic noise on the other hand becomes negligible with $>10^6$ sources.
            Finally, the two uncertainty models for photometry and spectroscopy we described above simulate only observational uncertainties. A discussion of systematic uncertainties is left to the final conclusions (Sect. \ref{conclusions}).
            
            \subsection{Description of templates} \label{templates}
            
            \begin{figure}
                \centering
                \includegraphics[width=0.5\textwidth]{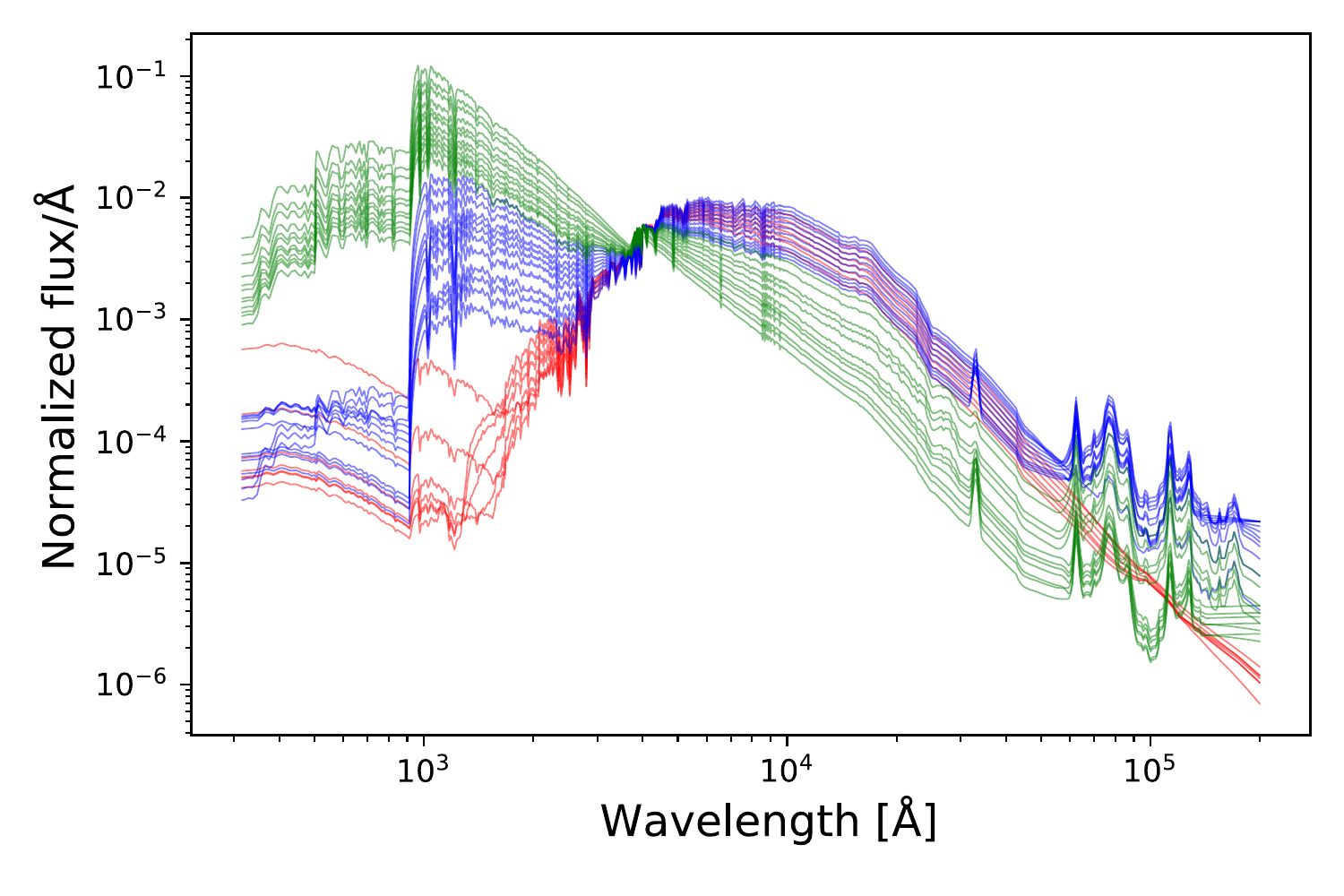}
                \caption{Arbitrary scaled COSMOS templates used for the Euclid Flagship mock galaxy catalogue. In red are shown the elliptical templates, in blue the lenticular and spiral and in green the starburst ones.}
                \label{fig:templates}
            \end{figure}
            For the ensemble photometric redshifts method to be successful, the analysis template set used to build the matrix $\mathbf{T}$ should be representative of the observed galaxy SEDs and also span the range of attenuation values.
            In this study we used the same template set that was used to generate the mock galaxy SEDs.
            Clearly this is an idealised situation that can lead to over-optimistic results.
            A discussion of this issue will be left to the final conclusions (Sect. \ref{conclusions}).
            The COSMOS template set includes a mix of elliptical, spiral, lenticular and starburst galaxy types making a total of $31$ templates \citep[indexed from $0$ to $30$;][]{2009ApJ...690.1236I}.
            The templates are plotted in Fig. \ref{fig:templates}.
            
            The three versions of the mock galaxy catalogue described in Sect. \ref{mock_data} used different assumptions on the attenuation and therefore required different template sets for analysis.
            In the first case, for the analysis of the non-attenuated catalogue we used templates without applying attenuation.
            This provided an idealised case study that we used to assess the impact that attenuation has on the result.
            
            In the second case with fixed attenuation the attenuation model is assumed to be known a priori. We applied the fixed attenuation model , which we described in Sect. \ref{mock_data}, to all  galaxy SEDs and also to the template set.
            
            Finally, in the third catalogue that is the most realistic case, realistic attenuation is applied to the mock galaxies. The attenuation curve and the $E(B-V)$ value are assigned to each galaxy from the Flagship catalogue.
            In this case we used a combination of attenuated and non-attenuated templates in the analysis.
            The attenuated templates were constructed by applying the attenuation factor (Eq. \ref{extinction_factor}) with the \citet{1984A&A...132..389P} and \citet{2000ApJ...533..682C} attenuation curves and $E(B-V)$ fixed to the median value from the Flagship catalogue, which cover a colour excess range from $0$ to $0.5$.
            \begin{table}
            \caption{Attenuation curves and COSMOS templates.}
                \centering
                \begin{tabular}{c c}
                \hline
                \\ [-1em]
                  COSMOS SEDs  & Attenuation curve \\ \hline
                  \\ [-1em]
                    0-9 & 0 \\
                    10-22 & 1 \\
                    23-30 & 2 \\
                    23-30 & 3 \\
                    23-30 & 4 \\
                    \hline
                \end{tabular}
                \vspace{1ex}

                {\raggedright \small \textbf{Notes.} The assignment of attenuation curves to the COSMOS templates. The attenuation curve identified by $0$ is a constant function per unit frequency, the index $1$ refers to the attenuation curve from \citet{1984A&A...132..389P}, the other attenuation curves, from $2$ to $4$, are from \citet{2000ApJ...533..682C}. \par}
                \label{tab:ext_SEDs}
            \end{table}
            In Table \ref{tab:ext_SEDs} we show the correspondence between the COSMOS templates and the attenuation curves. This procedure followed the recipe used for the Flagship mock galaxy catalogue (Carretero et al., private communication).
            In this way we had a very general set of $47$ templates that also contains the attenuation model. A more representative set of templates could be built using different values of $E(B-V)$; however, a greater number of templates would also increment the number of parameters that need to be fitted in order to compute the redshift distributions (see Sect. \ref{conclusions}).
            
            In order to build the matrix, the templates need to be shifted to $N_z$ redshifts (see Sect. \ref{machine_implementation}).
            The redshift distributions will be measured on the grid of these $N_z$ redshifts.
            We used the same redshift grid for the analysis of all the three catalogues, it ranged from redshift $0$ to $2.30$ with a step of $0.01$ for a total of $231$ redshifts.
            In the analysis of the real attenuation catalogue, the dimension of the template matrix was $(N_{\rm SED}\, N_z) \times (N_\lambda + N_{\rm b}) = 10\,857 \times 497$, while for the non-attenuated and fixed attenuation catalogue analyses it was $7161 \times 497$.
            
            \subsection{Colour selection} \label{sec:SOM}
            
            \begin{figure*}
                \centering
                \includegraphics[width=0.5\textwidth]{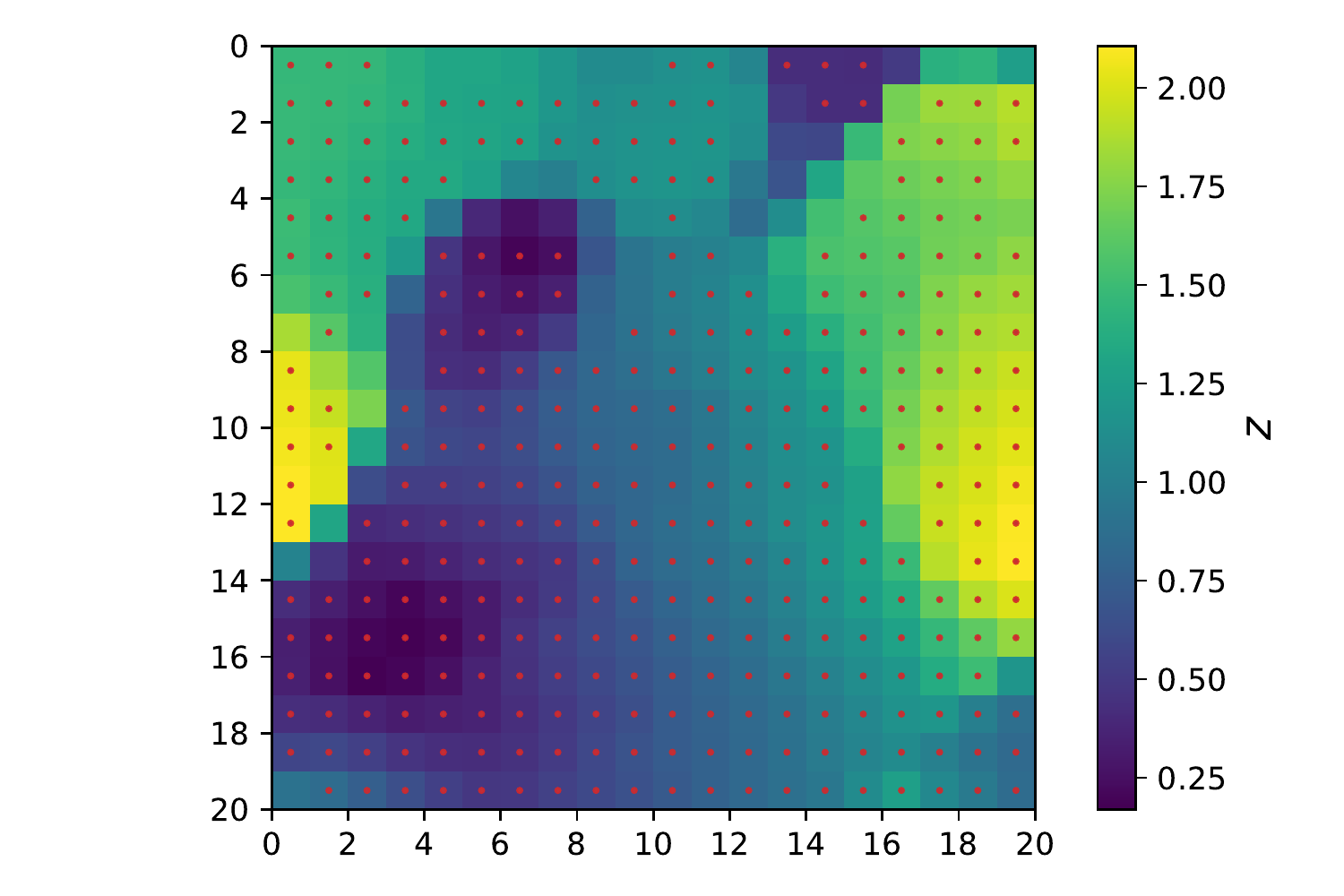}\includegraphics[width=0.5\textwidth]{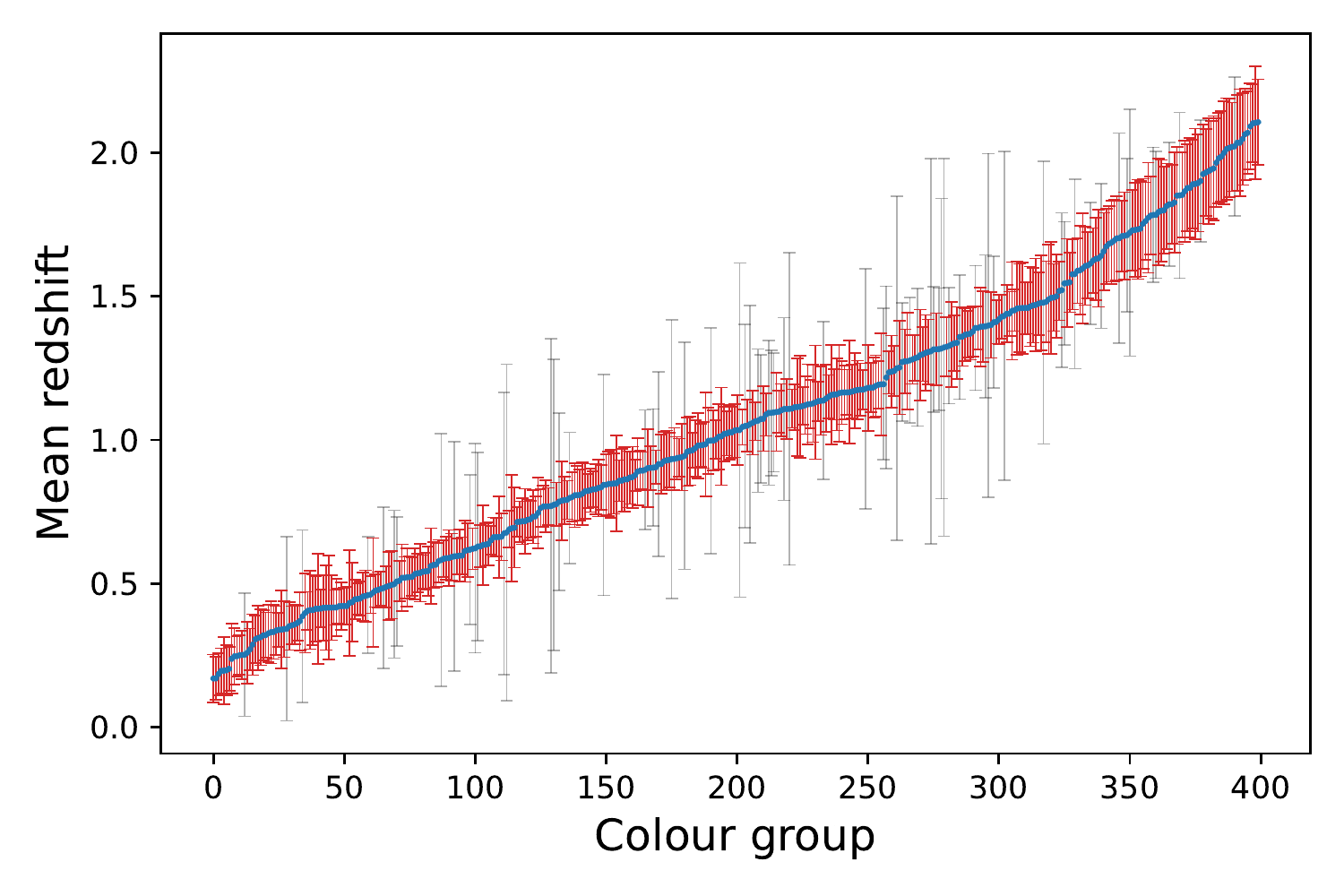}
                \caption{Photometric redshift division with SOM. \emph{Left}: the two-dimensional projection of the galaxy colour groups constructed with the SOM. The colour scale indicates the
                mean redshift of the SOM cells, which we define as colour groups. The red spots identify the cells with $\sigma_z < 0.20$. \emph{Right}: the colour groups sorted by their mean redshift. The error bar represents the standard deviation of redshift, $\sigma_z$, in each group. The groups with red error bars are the one with $\sigma_z < 0.20$.}
                \label{fig:som}
            \end{figure*}

            We used a self-organising map (SOM) for the colour group division.  The SOM \citep{kohonen-self-organized-formation-1982,SOM_1990} is an unsupervised machine learning algorithm that projects high-dimensional data on a lower-dimensional grid, usually a two-dimensional map.
            Its main characteristic is that the lower-dimension representation preserves the characteristic of the high-dimensional data.
            In the last few years SOMs have grown in popularity as a data driven method to estimate photometric redshifts \citep[e.g.][]{2015ApJ...813...53M,2020ApJ...888...83W}.
            However, in this work we simply exploited the efficiency with which SOMs are able to cluster and group data with similar features. Other approaches could be used to select the samples such as the K-means clustering algorithm, or photometric redshift bins.
            
            For each galaxy we had nine photometric fluxes (see Sect. \ref{mock_data}) from which we computed eight colours used to build the SOM.
            Using SOMPY, a SOM library for Python by \citet{moosavi2014sompy}, we built a $20 \times 20$ rectangular cell SOM using the principal component analysis (PCA) as the initialisation method.
            The SOMs we built have much smaller dimensions than the SOMs used for photometric redshift measurements or to estimate physical properties \citep{2015ApJ...813...53M,Davidzon2019}. We had two goals when we chose the SOM size: firstly, we wanted galaxy samples that spanned relatively narrow redshift ranges which are appropriate for measuring clustering statistics. Secondly, we needed these samples to be highly populated in order to be able to average out the spectroscopic noise during the stacking process. However, there is an intrinsic tension between these two goals due to the nature of the SOM algorithm. In this work we deemed more important the second goal and opted for a coarse SOM (see Appendix \ref{colours-in-som}) to have highly populated cells ($N_{\rm gal}>10^6 $) without the need to group cells together. Comparing Fig. \ref{fig:stack} left and right panels it can be seen how the increase in the number of galaxies in the cells reduces the noise in the stacked spectroscopy making it negligible; the issues related to the analysis of less populated groups (see Fig. \ref{fig:stack} left panel) will be discussed in Sect. \ref{conclusions}. Nevertheless, the choice of the SOM size will depend on the application, such as galaxy clustering studies or weak lensing, and should be optimised for the analysed survey. 

            We used the cells defined by the SOM grid to define the colour groups for analysis.
            In Fig. \ref{fig:som} we illustrate how the cells of the SOM grid map to redshift.
            In the analysis we focused on colour groups with compact redshift distributions and, therefore, we selected groups with standard deviation in redshift below 0.20 which are marked with red spots in in the left panel of Fig. \ref{fig:som}.
            The mean number of galaxies in a group is $2.5\times10^6$.

            \subsection{Quantifying the method performance}

            In order to quantify the error in the redshift distributions measured using the ensemble photometric redshift method and to understand if they can be useful in cosmological studies we compared the angular position of the BAO peak computed with the real redshift distribution and the one measured with the ensemble photometric redshifts method.
            
            The position of the BAO peak in the angular correlation function is determined by the projection of the sound horizon $r_{\rm s}$ at the comoving distance $r(z)$ to the galaxy sample. In the case of a thin redshift shell at redshift $z$, the angular scale is $\vartheta_{\rm BAO}= r_{\rm s}/r(z)$ \citep{Sanchez2011}. A systematic shift in the redshift distribution $\Delta_z$ propagates to an error in the angular position to first order as
            \begin{equation}
            \frac{\Delta \vartheta_{\mathrm{BAO}}}{\vartheta_{\mathrm{BAO}}} = \left. \frac{1}{r(\bar z)}\,\frac{\diff r}{\diff z}\right\rvert_{\bar z} \Delta_z\,,
            \label{D_BAO-D_z}
            \end{equation}
            where $\bar{z}$ is the mean redshift of the shell.
            
            However, the full shape of the angular correlation function also depends on the evolution of the correlation function integrated over the redshift distribution.
            This can have a substantial impact on the measurement of the BAO scale particularly when the redshift distribution has extended tails.
            We therefore used a full model of the angular correlation function to propagate the error. We wrote the angular correlation function in terms of the three-dimensional galaxy power spectrum $P_{\rm g}(k, z)$ and normalised redshift distribution $p(z)$,
            \begin{equation}
                w(\vartheta) = \int \frac{\diff \ell \, \ell}{2 \pi} \,J_0(\ell \vartheta) \int \diff r \,\frac{\left[ p(z)\, \frac{\diff z}{\diff r} \right]^2}{r^2} \, P_{\rm g} \left(\frac{\ell + \frac{1}{2}}{r}, z \right)\,,
            \label{correlation_function}
            \end{equation}
            where $r$ is the radial comoving distance, $J_0$ the Bessel function of order zero and the redshift $z=z(r)$ is a function of the radial comoving distance \citep{2018PhRvD..98d2006E}. This expression was derived using the Limber and flat-sky approximations $\left( k \longrightarrow \frac{\ell + 1/2}{r} \right)$ which are valid at the redshifts we probe. We modelled the relation between the galaxy power spectrum, $P_{\rm g}(k,z)$, and the matter power spectrum, $P_{\rm m}(k,z)$, with a linear bias $P_{\rm g}(k,z)=b^2 P_{\rm m}(k, z)$.
            
            We computed the matter power spectrum using the CLASS code \citep{2011JCAP...07..034B}.
            We adopted a flat $\mathrm{\Lambda CDM}$ cosmological model with $h  = 0.7$, $\Omega_{\rm b} = 0.05$, $\Omega_{\rm CDM}  = 0.25$ and $\Omega_{\mathrm{\Lambda}}  = 0.7$. We used only the position of the BAO peak in the analysis so the details of the galaxy bias model, overall power spectrum amplitude and broad-band shape of the power spectrum do not significantly influence the results.
            We integrated Eq. (\ref{correlation_function}) numerically over the redshift range $0<z<2.3$ (the limit of the mock catalogue).
            To achieve convergence the integration range was set with $k_{\rm max}=10$ which corresponds to $\ell_{\rm max}=\num{53000}$.

            To locate the BAO peak, we fitted the angular correlation function computed with Eq. (\ref{correlation_function}) with a template that consists of a power law, which describes the decreasing part of the correlation function, added to a Gaussian component that represents the peak,
            \begin{equation}
                f(\vartheta) = c_1\, \vartheta^{-\gamma} + c_2 \, \mathrm{exp}\left[ -\frac{(\vartheta - \vartheta_{\mathrm{BAO}})^2}{\sigma^2} \right] + y_{\rm norm}\,.
            \label{BAO_fit}
            \end{equation}
            There are six parameters in this model: $y_{\rm norm}$, an integration constant, $c_1$ and $c_2$, two coefficients and the three parameters characteristic of the correlation function, $\gamma$, which determines the slope, $\sigma$, which is the BAO peak width, and finally $\vartheta_{\mathrm{BAO}}$, the angular position of the BAO.
            We used the {\tt optimize.curve\_fit} algorithm from SciPy \citep{2020SciPy-NMeth} to fit the correlation function.
            We computed the BAO angular position using both the measured redshift distribution and the real one, known from the mock catalogue data (see Sect. \ref{mock_data}), for every colour group.
            The error in the measured redshift distribution was quantified by the relative error of the BAO angular position (hereafter BAO relative error):
            \begin{equation}
                \frac{\Delta \vartheta_{\mathrm{BAO}}}{\vartheta_{\mathrm{BAO}}}  = \frac{\vartheta_{\mathrm{BAO}}^{\rm real} - \vartheta_{\mathrm{BAO}}^{\rm fit}}{\vartheta_{\mathrm{BAO}}^{\rm real}}\,.
            \label{relative_error}
            \end{equation}
            
            We used the error on the BAO scale as the performance metric; however, the ensemble photometric redshift method may also be applied in tomographic weak lensing analyses. In the context of weak lensing, estimates of the mean redshifts of the tomographic bins are needed. Equation (\ref{D_BAO-D_z}) shows that in the limit of narrow redshift distributions, the error on the BAO scale can be equated with the error on the mean redshift. Therefore our results can also be interpreted as systematic errors for weak lensing analyses.


            \subsection{Optimising regression parameters}\label{parameter_selection}
            
            The regression methods LASSO and ElasticNet described in Sect. \ref{machine_implementation} have free parameters that must be chosen. We analysed the fixed attenuation catalogue over a range in the parameter
            space to test the quality of the regression result and tune the parameters. The LASSO method has the single parameter $\alpha_{\rm LASSO}$, while ElasticNet has two parameters $\alpha_{\rm ElasticNet}$, $\beta_{\rm ElasticNet}$. We quantified the goodness of fit with the BAO relative error, Eq. (\ref{relative_error}).
            
            We found that the regression parameters are weakly dependent on the mean redshift, $z_{\rm mean}$, of the analysed colour group.
            Therefore we decided to limit the redshift range of our analyses by selecting only
            colour groups with $z_{\rm mean} > 1$. Due to this selection we could neglect the regression parameters dependence on the redshift and use the same set of parameters for all the analysed groups. 
            As discussed in the introduction, we are most interested in the ability of the method to fit redshift distributions at high redshifts, $z > 1$, rather than low redshift ones.
            At $z>1$ the \Euclid near IR spectroscopy will measure the rest-frame
            SED at $\lambda < 9000 \, \AA$ which carries more information in the continuum shape to constrain photometric redshifts.
           
            \begin{figure}
                \centering
                \includegraphics[width=0.5\textwidth]{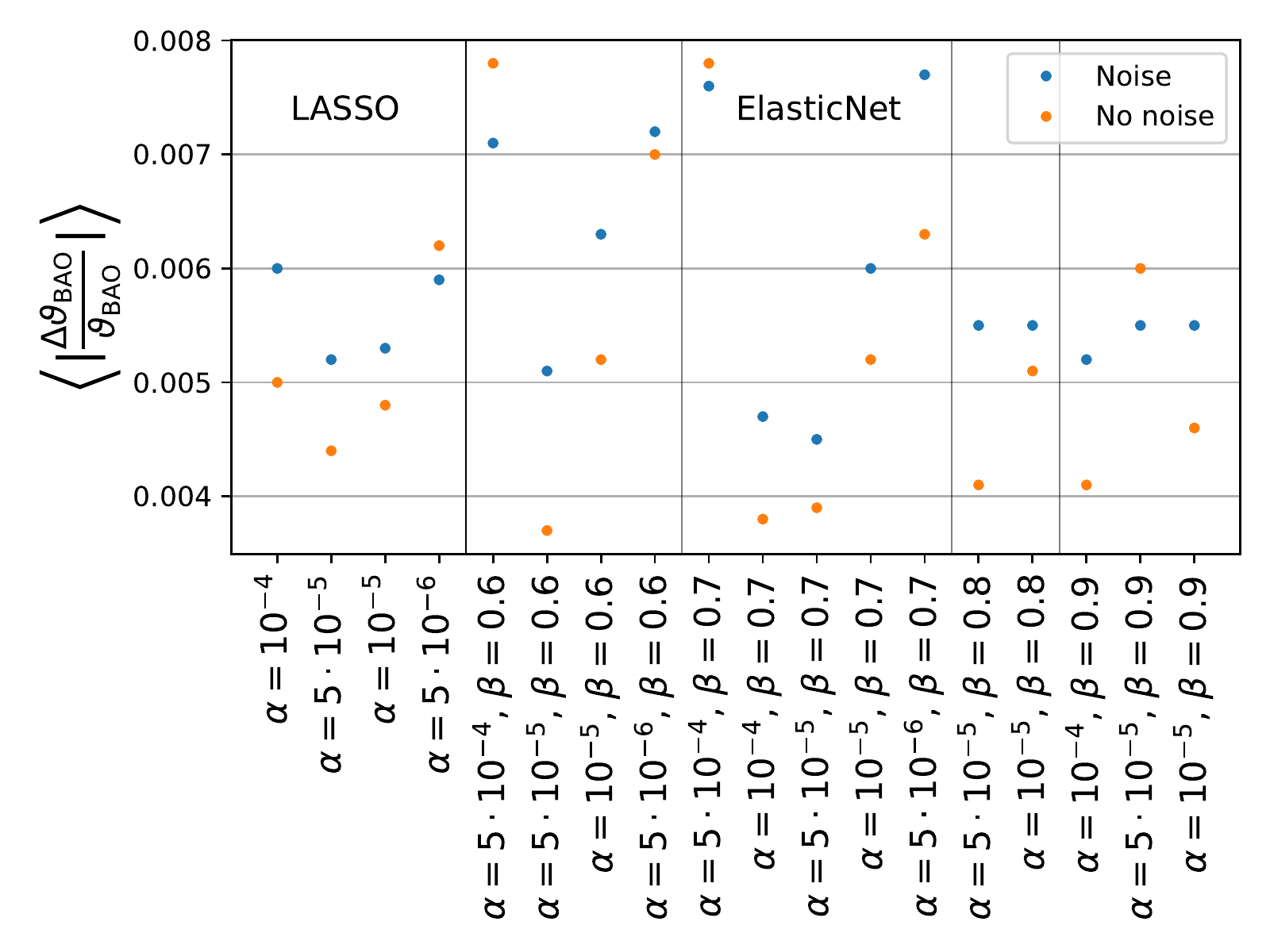}
                \caption{Mean BAO error for different sets of LASSO and ElasticNet fit parameters. The colour groups for the analysis were selected to have $z_{\rm mean} > 1$ and standard deviation in redshift lower than $0.15$.}
                \label{fig:fit_parameters}
            \end{figure}
            In Fig. \ref{fig:fit_parameters} we show the means over the colour groups of the absolute relative error on the BAO scale (hereafter mean BAO error) obtained with different parameters sets.
            For LASSO we observed a clear minimum in the goodness of fit that identifies $\alpha = 5 \times 10^{-5}$ as the best fit parameter both for the analyses with and without spectroscopic noise. 
            We used this value as $\alpha_{\rm LASSO}$ for all the analyses we present in the following section.
            On the other hand, for the ElasticNet method we found a broad minimum in the parameter space.
            The analyses without and with noise respectively have best fit parameters $\alpha_{\rm no \, noise} = 5 \times 10^{-5}$ and $\beta_{\rm no \, noise} = 0.6$, and $\alpha_{\rm noise} = 5 \times 10^{-5}$ and $\beta_{\rm noise} = 0.7$.
            These two sets of fit parameters were used as $\alpha_{\rm ElasticNet}$ and $\beta_{\rm ElasticNet}$ in the following section; we used the parameters with subscript `no noise' to compute the results presented in Sect. \ref{sec:no_noise}, and the ones with subscript `noise' for the analyses in Sect. \ref{sec:noise}.

            We found that the parameter choice also affects the smoothness of the redshift distribution estimates seen by eye. However, it was not possible to achieve the minimum error and smoothness simultaneously. We therefore optimised only for the error.
            
            \section{Results} \label{results}
            
            Having explained how the method is implemented, the data prepared and the redshift distribution computed we now discuss the results we obtained in the analyses of the three catalogues. We present the joint analysis of stacked photometry and spectroscopy. The analysis with  broad-band photometry only or spectroscopy only proves to be significantly less accurate. We discuss these cases in Appendix \ref{sp_vs_p}.
            
            In Sect. \ref{mock_data} we discussed how uncertainties are added to the photometry and the spectroscopy.
            The errors in the spectral fluxes are greater than those in photometric data (see Fig. \ref{fig:stack}) and we expect them to be the main sources of uncertainty in the redshift distribution fits; thus, we firstly analysed the catalogues adding only the photometric error and considered the spectroscopic noise only in later analyses.
            The analyses without spectroscopic noise can be considered as the limit in which there are enough galaxies in a colour group such that the noise in the stacked spectrum is negligible.
            
            \subsection{Analyses without noise} \label{sec:no_noise}
           
            \begin{figure*}
                \centering
                \includegraphics[width=0.5\textwidth]{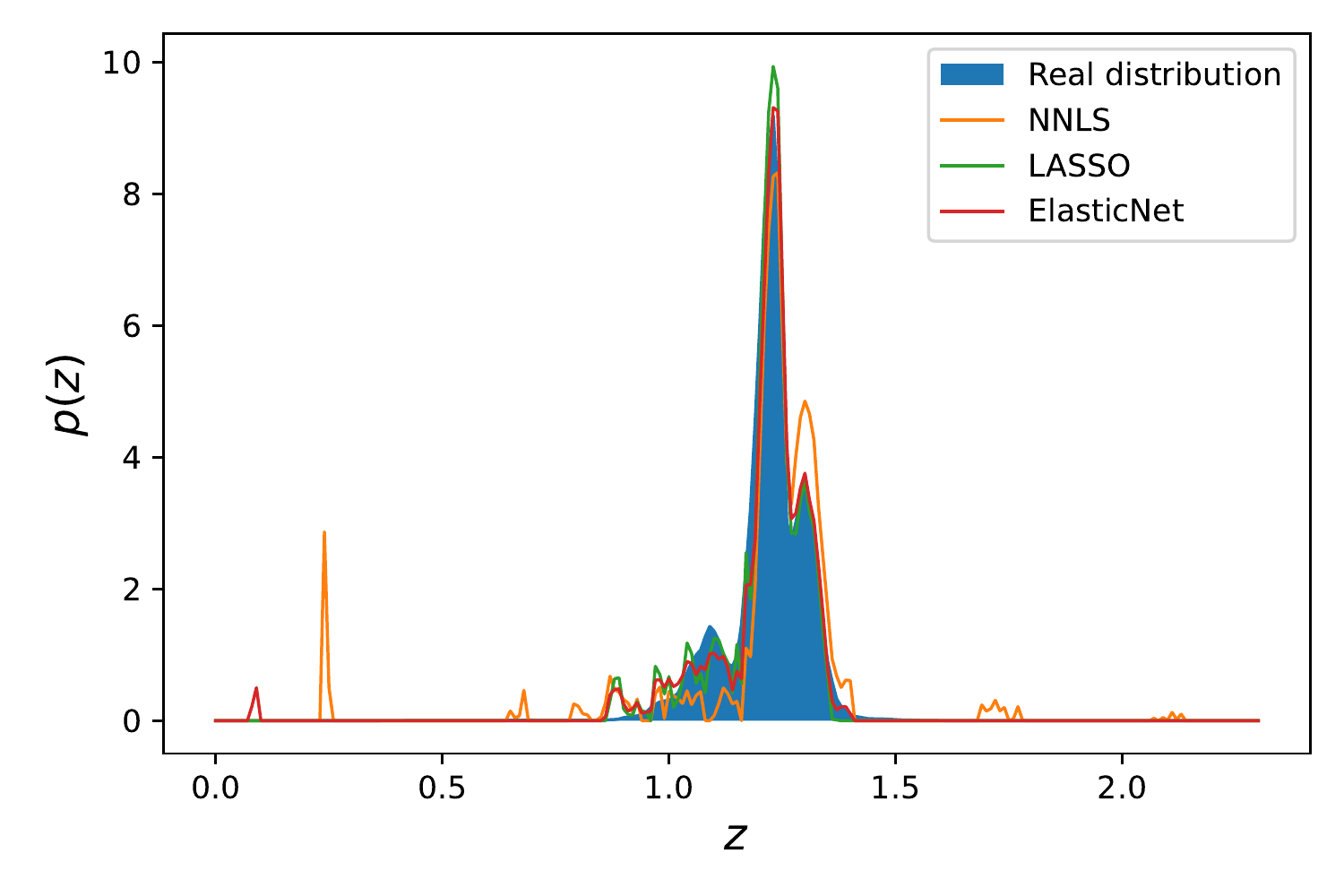}\includegraphics[width=0.5\textwidth]{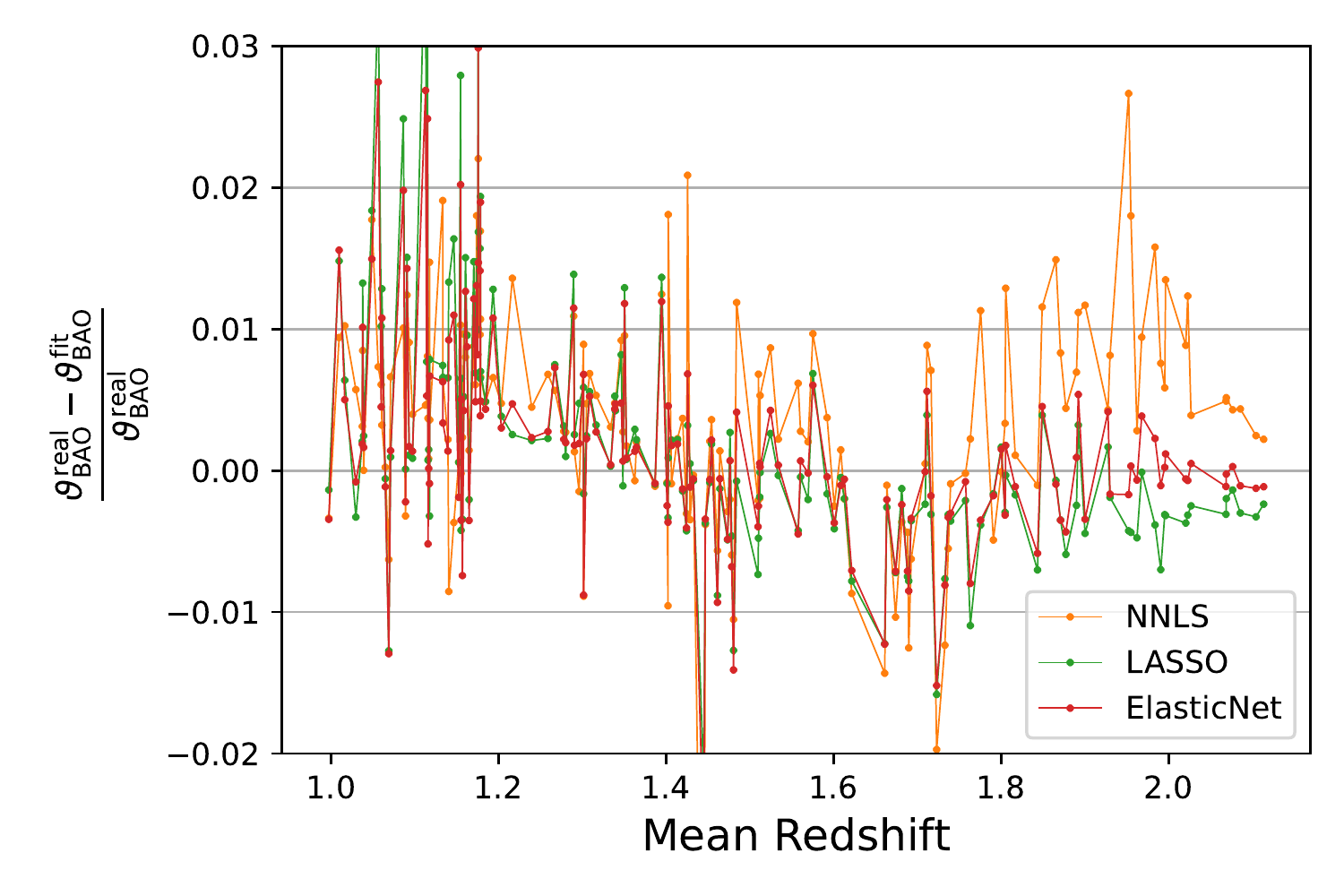}
                \includegraphics[width=0.5\textwidth]{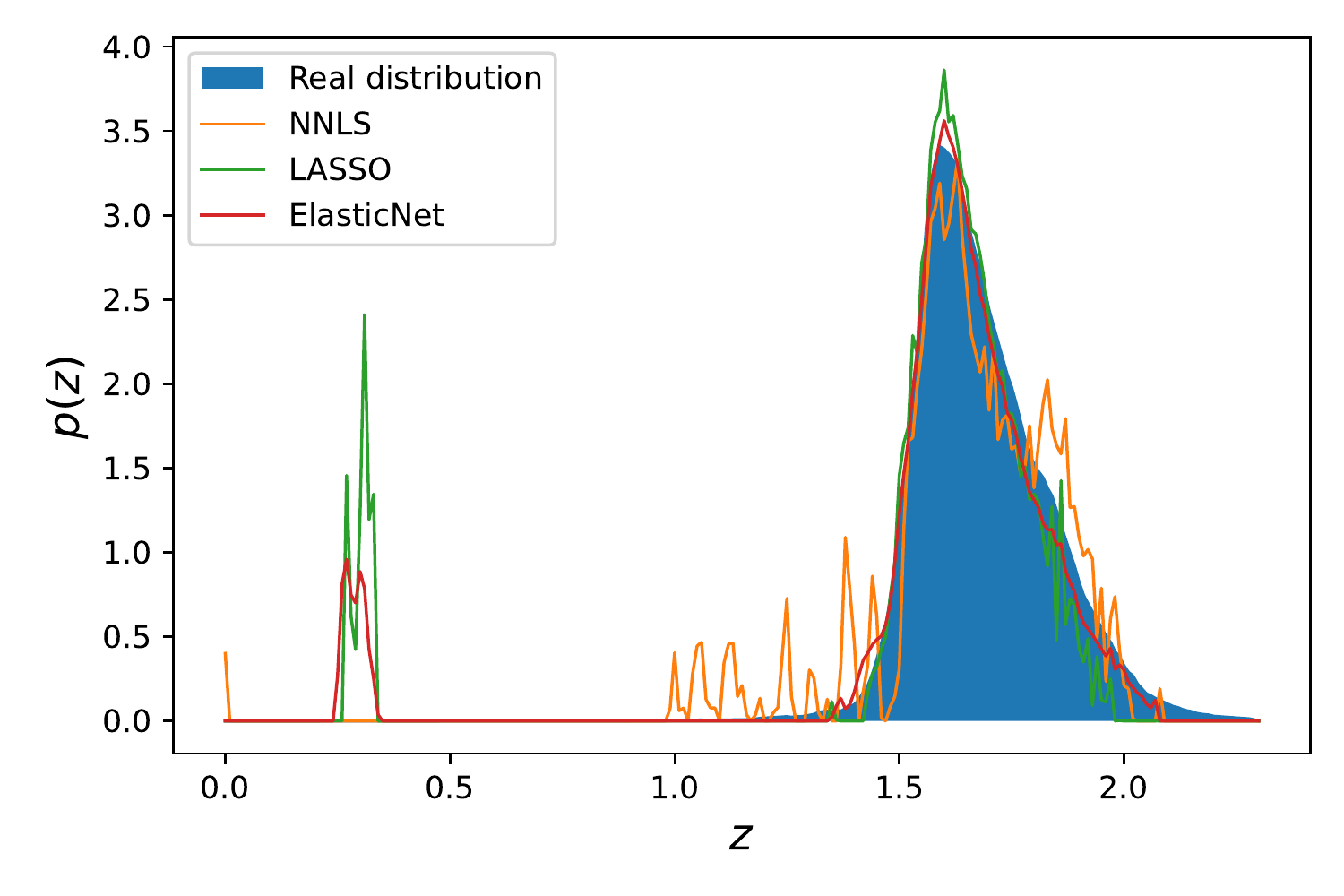}\includegraphics[width=0.5\textwidth]{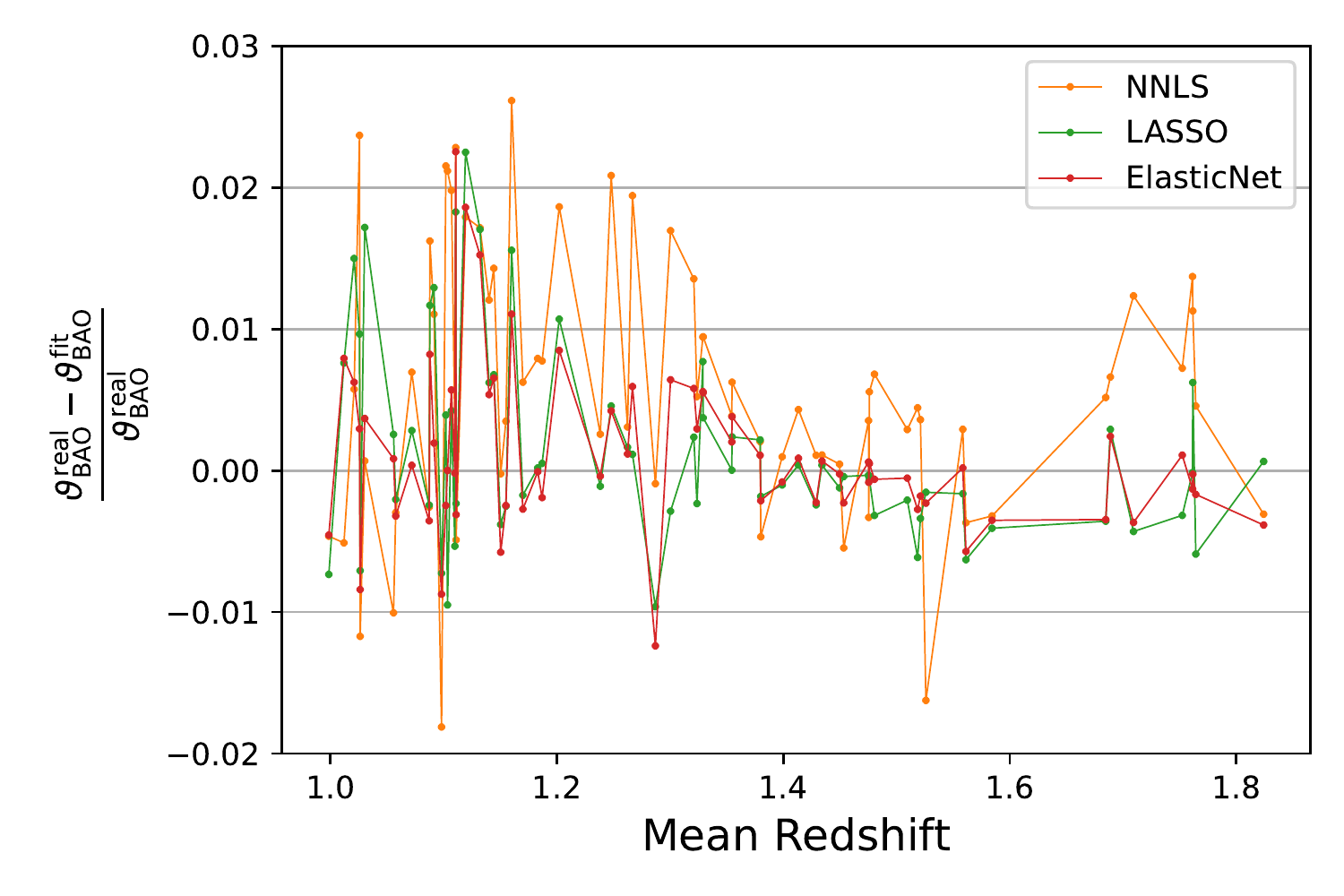}
                \includegraphics[width=0.5\textwidth]{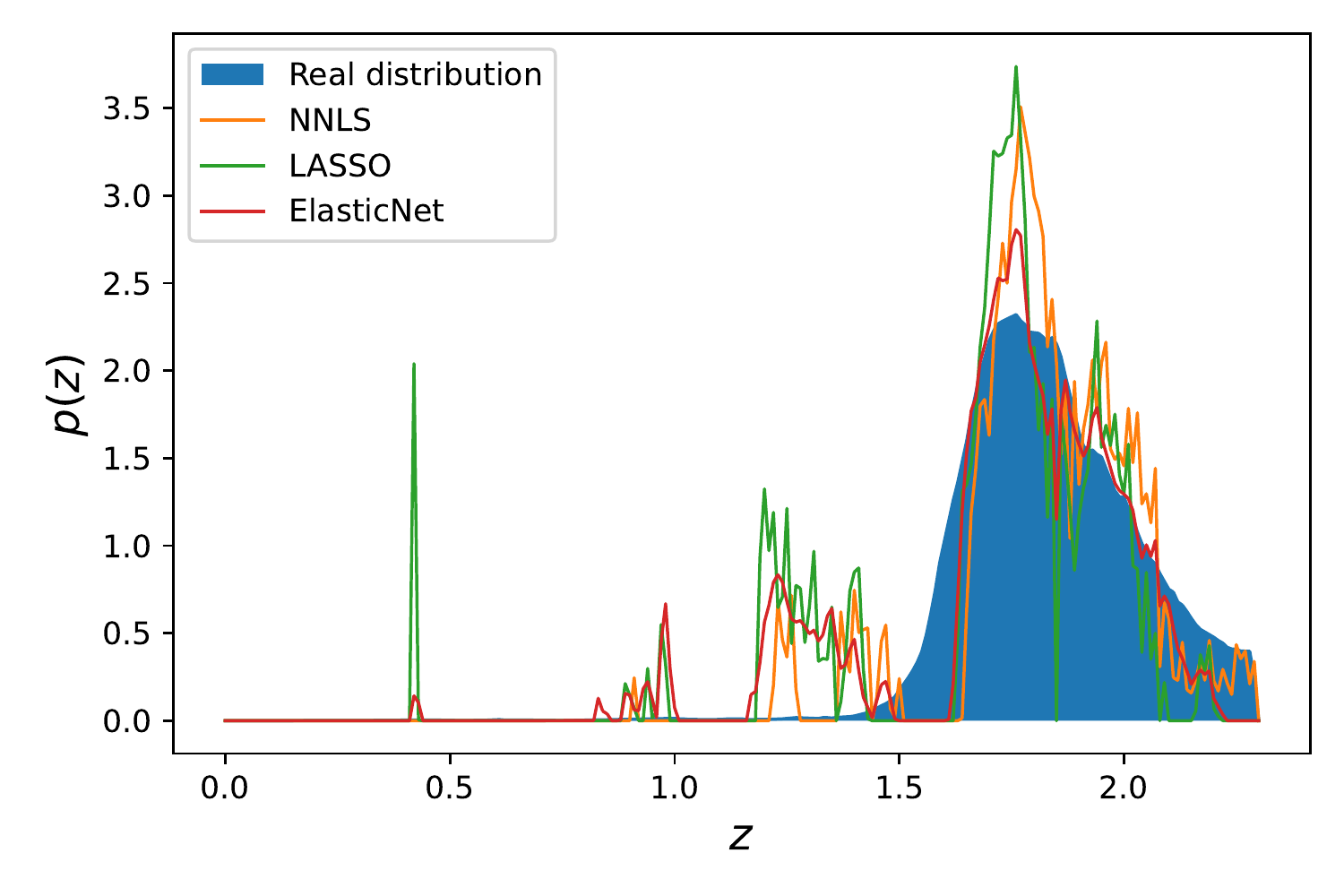}\includegraphics[width=0.5\textwidth]{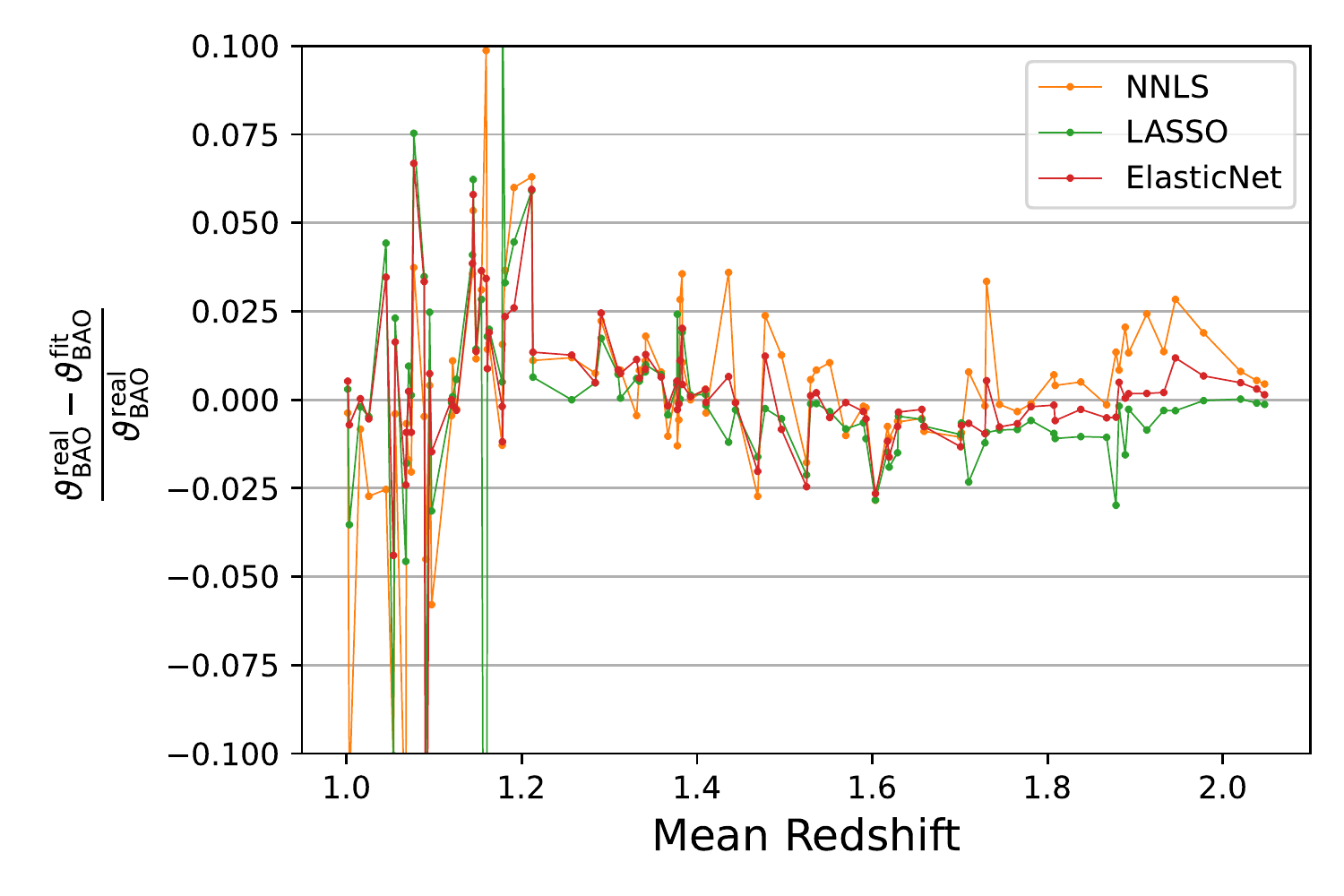}
                \caption{Results on ideal spectroscopy without measurement noise. \emph{Top row}: results of the non-attenuated catalogue analyses. \emph{Middle row}: results of the fixed attenuation catalogue analyses. \emph{Bottom row}: results of the real attenuation catalogue analyses. NNLS, LASSO and ElasticNet results are respectively plotted in orange, green and red. The left column shows examples of redshift distribution fits with the real distribution plotted in blue. Shown on the right is the BAO relative error of the analysed colour groups ordered by redshift.}
                \label{fig:no_noise}
            \end{figure*}
            
            \begin{table}
            \caption{Results.}
                \centering
                \begin{tabular}{cccc} \hline
                     \\ [-1em]
                     \multicolumn{4}{c}{Analysis without spectroscopic noise} \\ [0.1ex]
                     \hline
                     \\ [-1em]
                     & NNLS & LASSO & ElasticNet \\ \hline
                     \\ [-1em]
                     non-attenuated & 0.0069 & 0.0062 & 0.0053\\
                     fixed & 0.0086 & 0.0050 & 0.0040\\
                     real & 0.019 & 0.021 & 0.014\\ 
                     \hline \hline
                     \\ [-1em]
                     \multicolumn{4}{c}{Analysis with spectroscopic noise} \\ [0.1ex]
                     \hline
                     \\ [-1em]
                     & NNLS & LASSO & ElasticNet \\ \hline
                     \\ [-1em]
                     non-attenuated & 0.0073 & 0.0071 & 0.0063\\
                     fixed & 0.0099 & 0.0066 & 0.0057\\
                     real & 0.016 & 0.023 & 0.015\\ 
                     \hline
                \end{tabular}
                \vspace{1ex}

                {\raggedright \small \textbf{Notes.} Mean BAO errors for the analyses without and with spectroscopic noise. \par}
                \label{tab:BAO_error}
            \end{table}
            
            In the left panels of Fig. \ref{fig:no_noise} we present an example of a fitted redshift distribution for each one of the analysed catalogues.
            With these plots we highlight not only the general features related to the analysis without spectroscopic noise, but also the characteristics of the different catalogue analyses.
            
            Firstly, in the non-attenuated and fixed attenuation analyses the method is able to recover and fit the position, the width and the height of the redshift distributions; moreover, the detailed shapes of the distributions are well fitted. The three algorithms, NNLS, LASSO and ElasticNet give comparable results.
            However, in some cases spurious secondary peaks are evident. The occurrence of spurious peaks is reduced in the LASSO and ElasticNet results due to the shrinkage and selection processes in the algorithms which suppress  secondary solutions in favour of the principal ones.
            
            Secondly, we find a loss of precision in the real attenuation case. All three estimators fail to reproduce the shape of the redshift distributions accurately. As seen in the example shown in Fig. \ref{fig:no_noise}, the principal peak in the estimated redshift distribution is narrower and many significant spurious peaks are seen.
            
            The right-hand panels of Fig. \ref{fig:no_noise} show the relative error in the recovered BAO position, Eq. (\ref{relative_error}), for each colour group. The error is weakly dependent on the mean redshift of the group. The three estimators tend to show more stable performance and give lower error at higher redshift $z>1.4$. The NNLS estimator gives the largest error while the LASSO and ElasticNet estimators perform similarly.
            
            The best performance is found in the case of the non-attenuated catalogue
            which has a mean error of approximately 0.5--0.7\% in the BAO position from the three estimators (see Table \ref{tab:BAO_error}). The fixed attenuation analysis shows similar error of 0.4--0.9\%. However, when allowing the attenuation to be free in the real attenuation case, the error grows to 2\%. Attenuation introduces a degeneracy in colour-redshift space that leads to spurious peaks in the redshift distributions which biases the BAO angular position. Underestimation of the BAO position signifies that the redshift was biased high, as seen for the NNLS estimator at $z>1.8$.
            Overall the three methods have a similar error around $2\%$ in the presence of attenuation with NNLS and ElasticNet giving the best fits. LASSO on average performed less well due to a small number of groups that were poorly fitted. At $z>1.5$ the LASSO and ElasticNet algorithms perform better in the presence of attenuation, which may be attributed to the training process. Indeed, the performance will depend on the internal galaxy attenuation models used in the template set and the training sample.
            
            \subsection{Analyses with noise} \label{sec:noise}
            
            \begin{figure*}
                 \centering
                \includegraphics[width=0.5\textwidth]{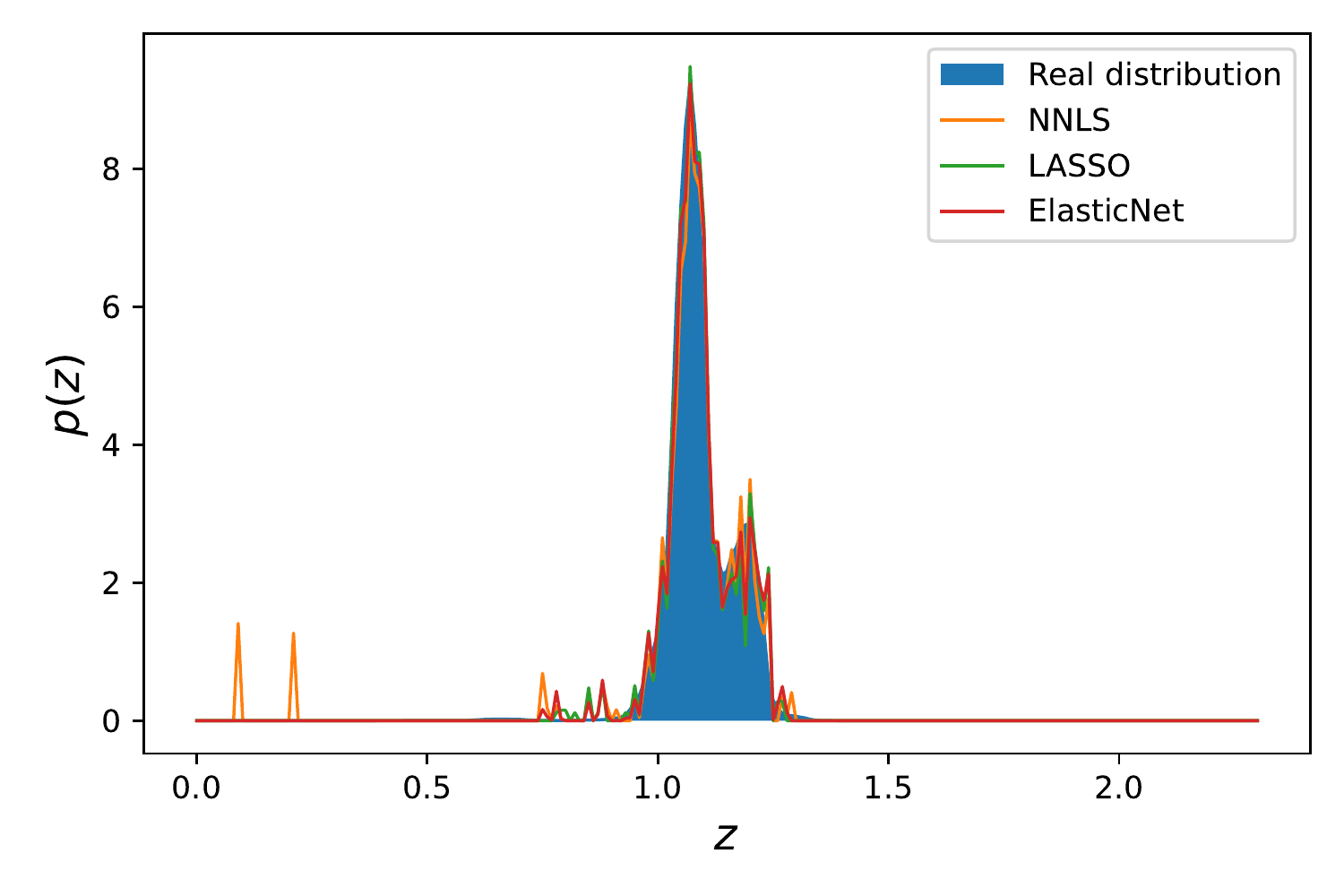}\includegraphics[width=0.5\textwidth]{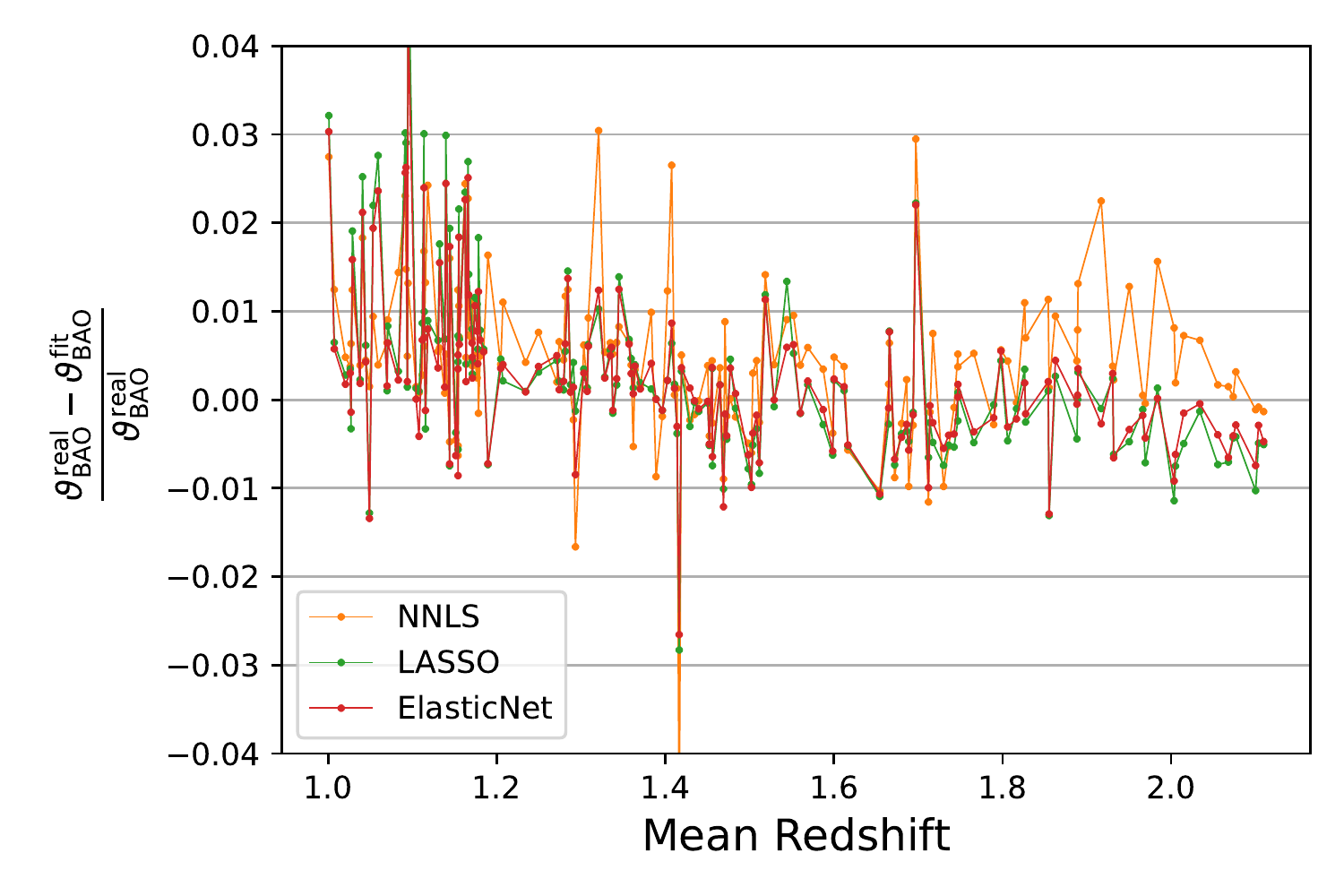}
                \includegraphics[width=0.5\textwidth]{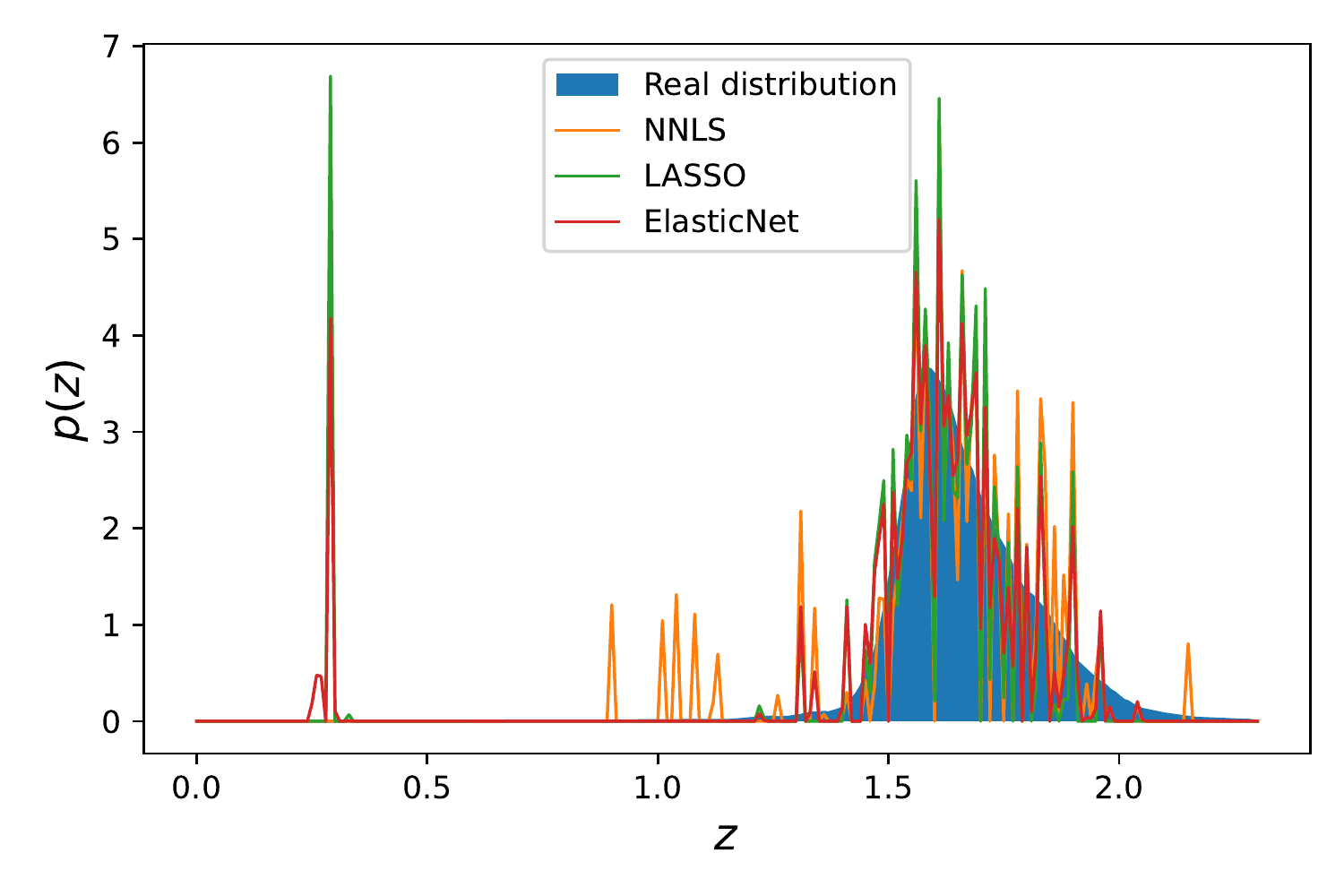}\includegraphics[width=0.5\textwidth]{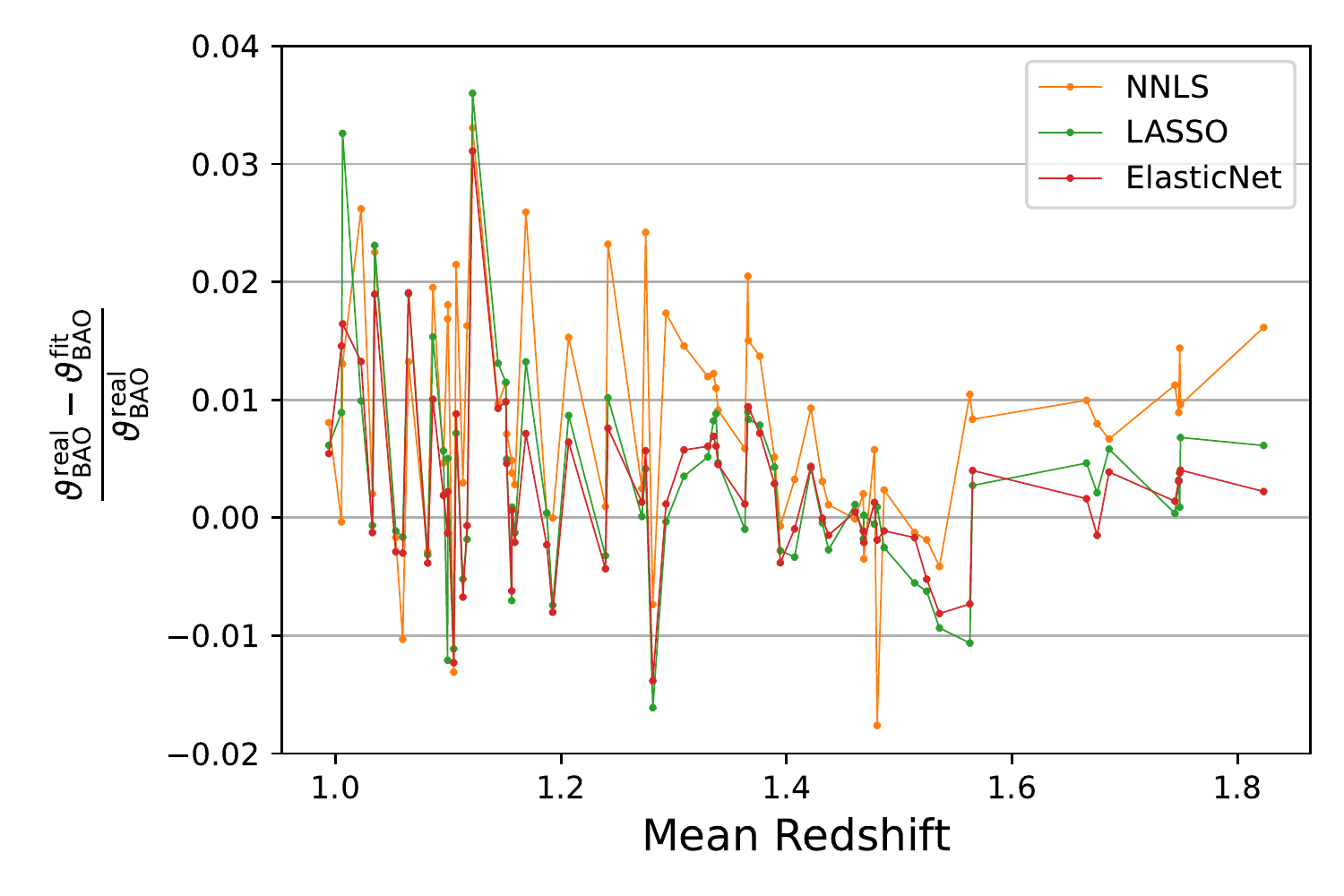}
                \includegraphics[width=0.5\textwidth]{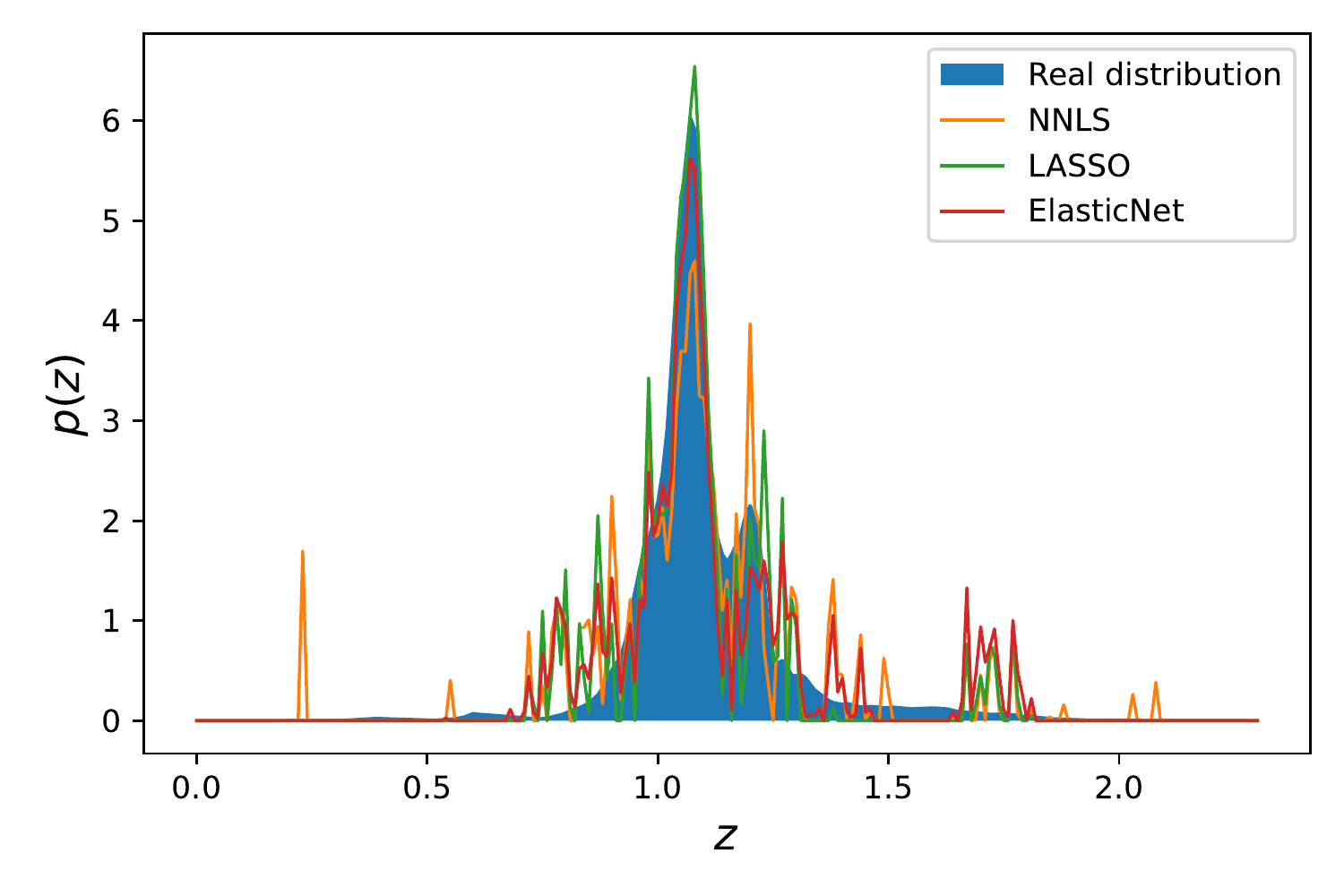}\includegraphics[width=0.5\textwidth]{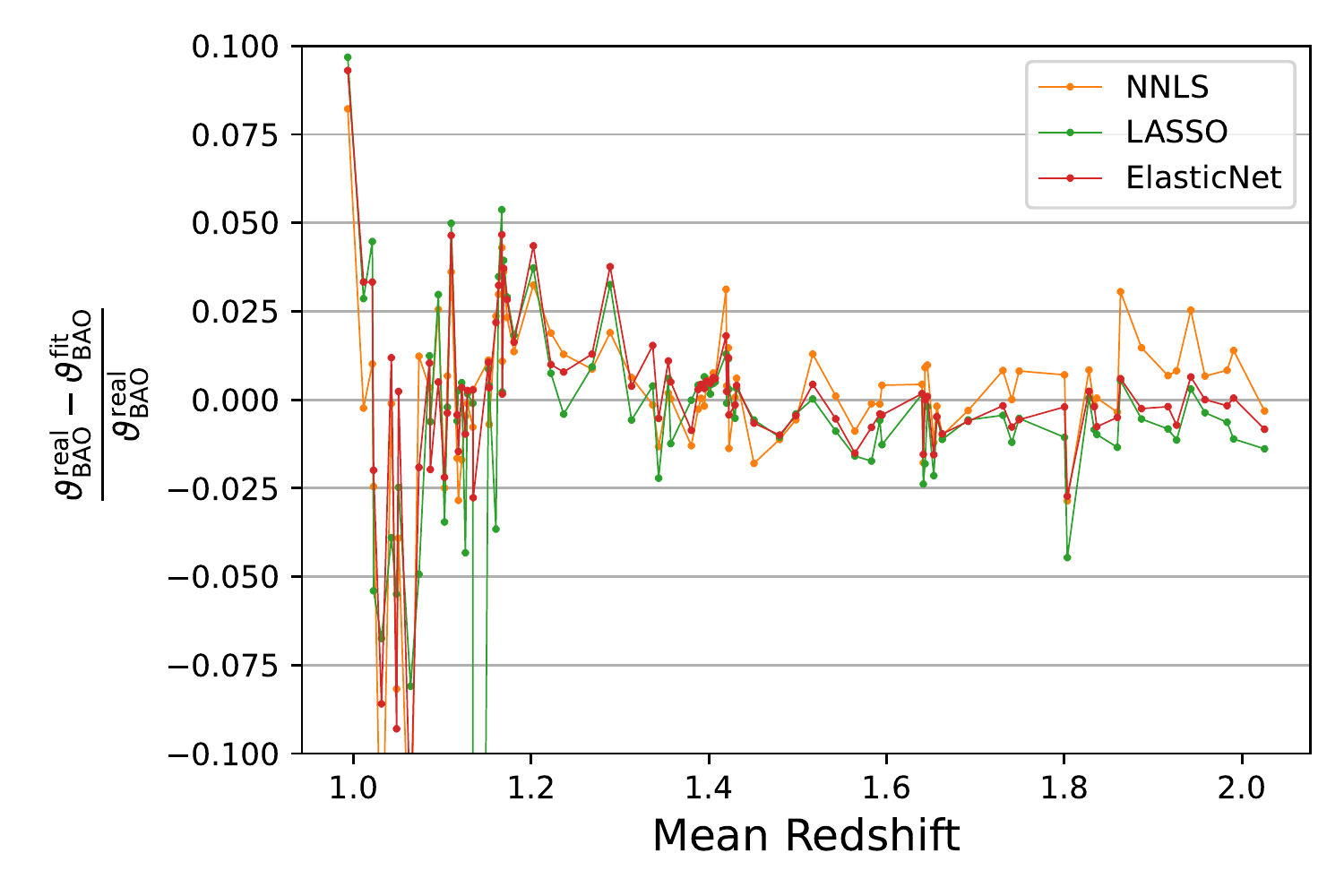}
                \caption{Results with noisy spectroscopy. 
                The panels show the same as in Fig. \ref{fig:no_noise}, the example redshift distribution fits plotted here are from different colour groups than the ones shown in Fig. \ref{fig:no_noise}. 
                The redshift distributions estimated for colour groups at higher redshift tend to be less smooth with a greater frequency of spurious peaks. The LASSO and ElasticNet regression algorithms have free parameters that can be adjusted to give smoother distributions but at the cost of lower accuracy.}
                \label{fig:noise}
            \end{figure*}
          
            The results of the analyses with noisy spectroscopy are shown in Fig. \ref{fig:noise}, on the left side panels we show an example redshift distribution for each one of the analysed catalogues. 
            The non-attenuated and fixed attenuation catalogue analyses with noise show similar results to the analyses without noise.
            At lower redshift the method is able to recover the redshift distributions with great detail with all three the regression methods, however that is not the case at higher redshifts, $z > 1.5$.
            At high redshift the computed distribution tend to be very noisy and less smooth (see Fig. \ref{fig:noise} middle left panel).
            The degree of this behaviour depends on the regression method used in the fit with NNLS showing more spurious peaks.
            Even though the smoothness of the redshift distributions is lost at high redshifts, the position and the width of the distributions are still recovered with enough precision in order to have small relative errors in the BAO angular position measurements. As for the analyses without spectroscopic noise we observe more stable performance and lower error at higher redshifts.
            
            In the real attenuation catalogue analysis we observe more spurious peaks separated from the principal peak than the ones in the other two catalogue analyses.
            However these spurious structures usually are more peaked and frequent with respect to the much wider ones we have in the noiseless analysis of this same catalogue.
            Moreover, the fitted redshift distributions tend to be very noisy and lose smoothness even at lower redshifts (see Fig. \ref{fig:noise} bottom left panel).
            
            The measured error on the BAO position are plotted in the right-hand side panels of Fig. \ref{fig:noise}. The non-attenuated and fixed
            attenuation cases show similar trends to the analyses without noise.
            In both cases we are able to recover the BAO position well, however, at $z>1.5$ the NNLS algorithm shows a trend of underestimating the BAO scale.
            
            In the case of the real attenuation analysis, colour groups at $z\sim1.05$ tend to have the BAO position overestimated. This trend was not evident in the analysis without noise and indicates an added degeneracy related to the attenuation curves and how attenuation is modelled in the presence of spectroscopic noise. At $z>1.5$ similar behaviour is found with and without spectroscopic noise.
            
            The mean BAO errors are reported in Table \ref{tab:BAO_error}. The trends are consistent with the analysis without spectroscopic noise but we find that the error degrades by approximately $10$\%. This indicates that the error introduced by the internal galaxy attenuation and its imperfect modelling is the most important factor that limits the fit performance.
            
            \section{Conclusions} \label{conclusions}
            In this pilot study we have tested the use of stacked spectra from \Euclid near infra-red grism spectroscopy to reconstruct the ensemble  redshift distribution of photometrically selected galaxy samples. The general approach in the context of slitless spectroscopic surveys was proposed by \citet{2019arXiv190301571P}. Here we considered the combination of broad-band photometry including the $ugrizy$ bands from the Vera C. Rubin Observatory and \Euclid NISP $YJH$ augmented with stacked NISP grism spectroscopy using the Euclid Flagship mock galaxy catalogue. Since the optimisation of the photometric galaxy selection in \Euclid is ongoing \citep{2021arXiv210405698E}, we selected mock galaxy samples in colour space using the SOM algorithm. These galaxy samples have compact distributions in both colour and redshift. The redshift distributions inferred from broad-band photometry alone prove to be unreliable as shown in Appendix \ref{sp_vs_p}. This is not unexpected since the constraints from SED fitting depend on the template priors which we do not consider \citep{Benitez2000}.  However, we find that the full application of the joint analysis of photometry and spectroscopy on mock survey data is promising and very informative of both the method's limits and its potential applications.
            
            To assess the quality of the redshift distribution estimation we focused on the cosmological application of inferring the BAO scale with photometric galaxy clustering measurements. Currently the best constraints of the BAO scale with photometric measurements is $\sim4\%$ \citep{Seo2012,DES_BAO}. This error depends on the survey area, the redshift of the sample as well as the width of the redshift distribution.
            We can expect that \Euclid will make measurements of the BAO scale with percent-level statistical precision in multiple redshift bins from $0<z<3$.
            Thus, it will be necessary to reduce the systematic error propagated from uncertainty in the redshift distributions to the sub-percent level.
            
            We tested the quality of the redshift distribution estimates in progressively more realistic cases on mock galaxy catalogues considering grism spectroscopy with and without measurement noise. In the most idealised configuration without internal galactic attenuation the redshift distributions were reconstructed with excellent accuracy on the BAO scale of about $0.5$\%. The presence of spectroscopic noise degraded this error to about $0.6$\%. We compared three regression algorithms, NNLS, LASSO and ElasticNet. All three performed well but ElasticNet which has two free parameters gave the best results.
             
            Our main conclusion is that the accuracy of the redshift distribution estimation is  limited primarily by internal galaxy attenuation and its modelling. Compared with the non-attenuated and fixed attenuation cases, we found a significant loss of precision in the real attenuation analysis where the attenuation curve varies for each galaxy. This was the case in both analyses we carried out considering spectra with and without measurement noise (Sects. \ref{sec:no_noise} and \ref{sec:noise}).  Nevertheless, despite the degeneracies introduced by attenuation we found that the BAO scale could be recovered with a precision better than 2\%.
            However, this behaviour reveals the importance of the template set and attenuation model that must be representative of the galaxy sample.
            
            On this matter, we made a preliminary investigation  to understand if increasing the number of templates affects the method performance. 
            We expanded the template set with three values of $E(B-V)$ to give three times the number of templates.  Using this template set, we analysed a random sub-sample of the  real attenuation catalogue without spectroscopic noise. The resulting error on the BAO scale  remained stable and did not decrease by the addition of more templates to the fit. So we expect that choosing a larger, but more representative set of templates for the fit will improve the method performance.
            
            In this work we used the same template set to build the galaxy spectra and the template matrix (see Sect. \ref{templates}); this is an ideal situation that is not possible when analysing real observations. We expect a further loss in precision in a realistic case when the template set is not fully representative of the galaxy sample. However, optimisations may be made in the template set with the addition of priors that may improve the fitting performance.
            Spectroscopic campaigns such as the ongoing C3R2 will build representative redshift catalogs that can provide invaluable information to improve the templates, constrain the attenuation models, and set priors.
            
            Another idealisation made while building the template matrix that needs to be highlighted is the range of the redshift grid. The redshift grid we used for our analyses covers only the redshift range that is simulated in the Euclid Flagship catalogue. Real catalogues will contain higher redshift galaxies, hence a wider range should be spanned by the redshift grid. Nevertheless, we still expect that extending the redshift range will not produce a significant loss in precision although we may find spurious peaks at high redshift if they are degenerate with the adopted attenuation model.
            
            Moreover, it may be possible to improve the method fitting performance by introducing inverse-error weights in the stacked spectrum and template matrix as we suggested in Sect. \ref{combining}. These weights will be useful in the analyses with noisy spectra to balance the relative importance of the photometry and spectroscopy in the fit and produce smoother redshift distributions. In addition it could make feasible the analysis of less populous colour groups, in which there are not enough galaxies to average the noise out of the stacked spectrum.
            
            In the case of real observations we should also account for contamination from stars and quasars for which the template fits may be unreliable. We will also face additional sources of systematic error that we have not addressed here. Grism spectroscopy suffers from contamination due to overlapping spectra \citep{2009PASP..121...59K}. This contamination can particularly spoil the measurement of the galaxy continuum. However, we expect that the spurious signals will be uncorrelated between spectra and average out in the stack. The spectrophotometric calibration error on the other hand can systematically alter the shape of all spectra in the stack and  bias the fit. The importance of these sources of error will be investigated in a later work.
            Future work must also investigate the effect of emission lines in the galaxy SEDs on the stacked spectrum. We expect emission lines to appear in the stacked spectrum as bumps, the width of which will depend on the photometric redshift bin width. In order to take the emission lines into account in the analysis, they need to be modelled and added to the template matrix SEDs. Potentially the emission lines signal would help constrain the template fitting and improve the results, but they could also make the analysis more sensitive to the choice of the template set.
  
            Our analyses confirm that in the case of \Euclid, stacked spectroscopy adds information that can help to break degeneracies in colour space that affect statistical studies based on photometric redshifts. The approach provides an internal method for calibrating the redshift distributions without relying on representative spectroscopic samples. This is particularly important at the high redshifts and faint galaxy luminosities probed by \Euclid where statistically complete samples of spectroscopic galaxy redshifts are lacking for calibration. 

            \begin{acknowledgements}
            
            \AckEC
            
            This work has made use of CosmoHub. CosmoHub has been developed by the Port d'Informació Científica (PIC), maintained through a collaboration of the Institut de Física d'Altes Energies (IFAE) and the Centro de Investigaciones Energéticas, Medioambientales y Tecnológicas (CIEMAT) and the Institute of Space Sciences (CSIC \& IEEC), and was partially funded by the "Plan Estatal de Investigación Científica y Técnica y de Innovación" programme of the Spanish government.
            \end{acknowledgements}
            
            \bibliographystyle{aa_url}
            \bibliography{paper_main}
            
            \begin{appendix}
            \section{Spectro-photometry vs. photometry} \label{sp_vs_p}
            
            \begin{figure}
                \centering
                \includegraphics[width=0.5\textwidth]{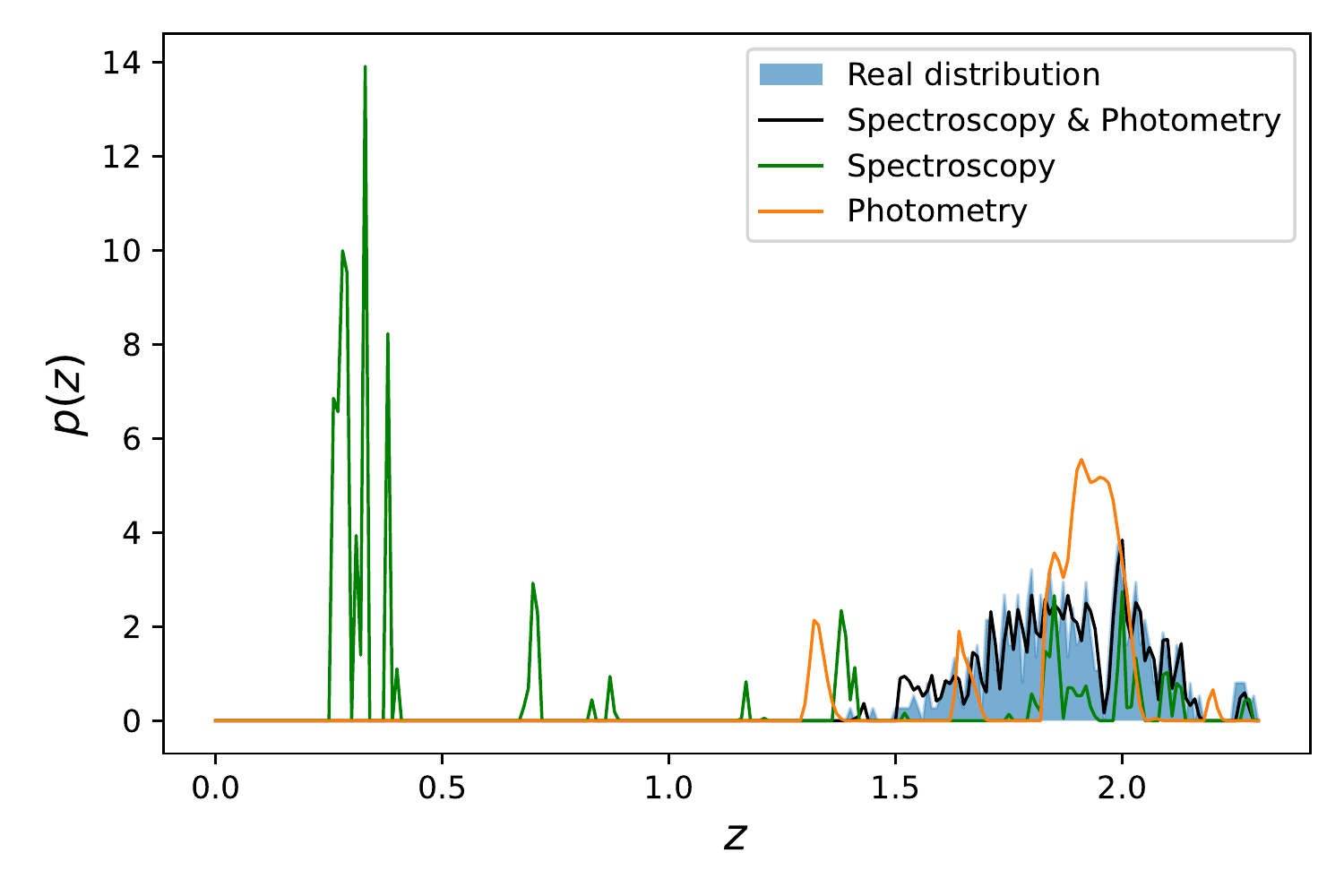}
                \includegraphics[width=0.5\textwidth]{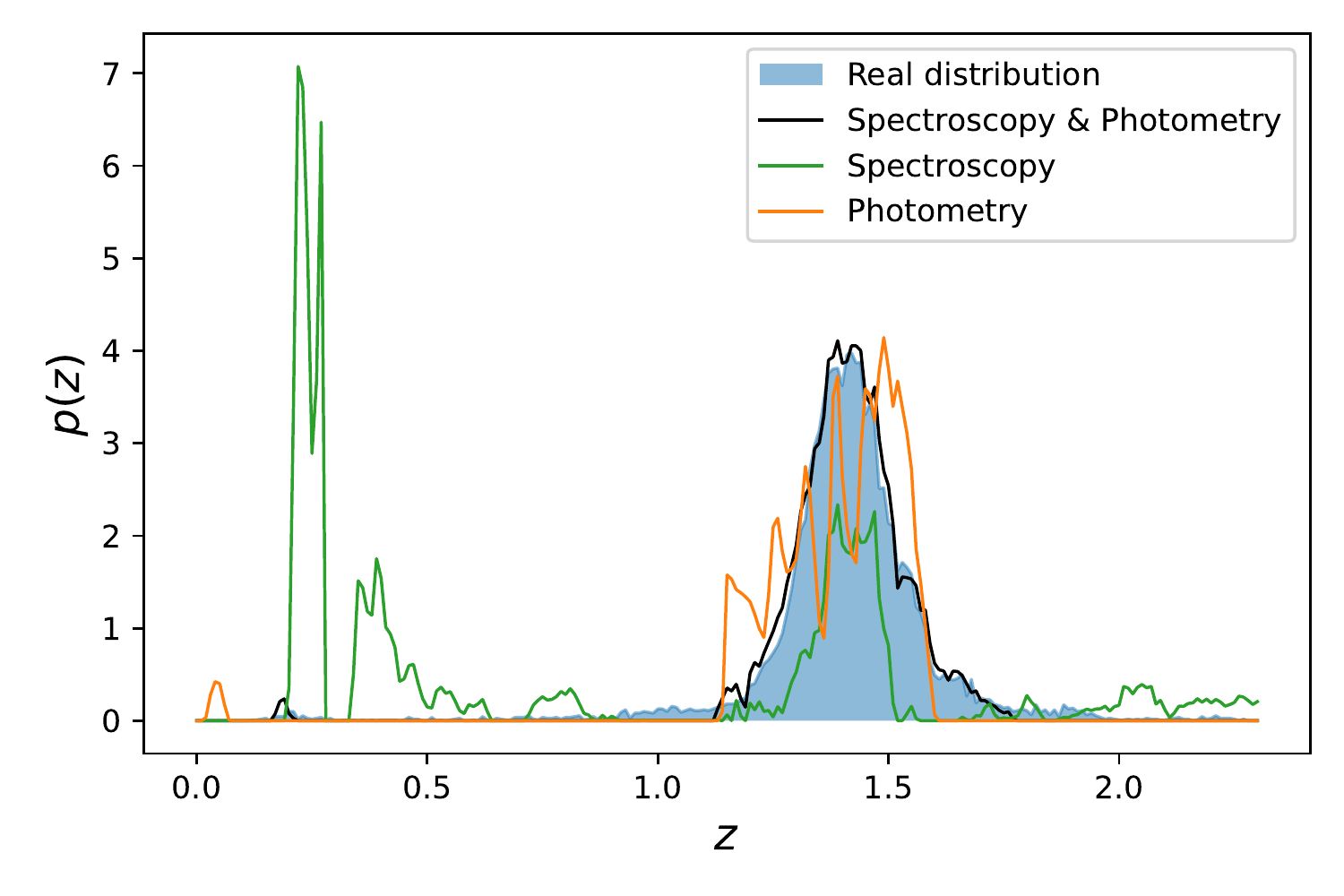}
                \caption{Two example redshift distributions from the analyses of the non-attenuated catalogue, top panel, and the real attenuation one, bottom panel, without the measurement noise and obtained with the ElasticNet regularisation. The results from the combination of stacked spectroscopy and photometry, stacked spectroscopy alone and stacked photometry are respectively plotted in black, green and orange. The real distribution is the filled blue histogram.}
                \label{fig:comparison-pz}
            \end{figure}
            \begin{figure}
                \centering
                 \includegraphics[width = 0.5\textwidth]{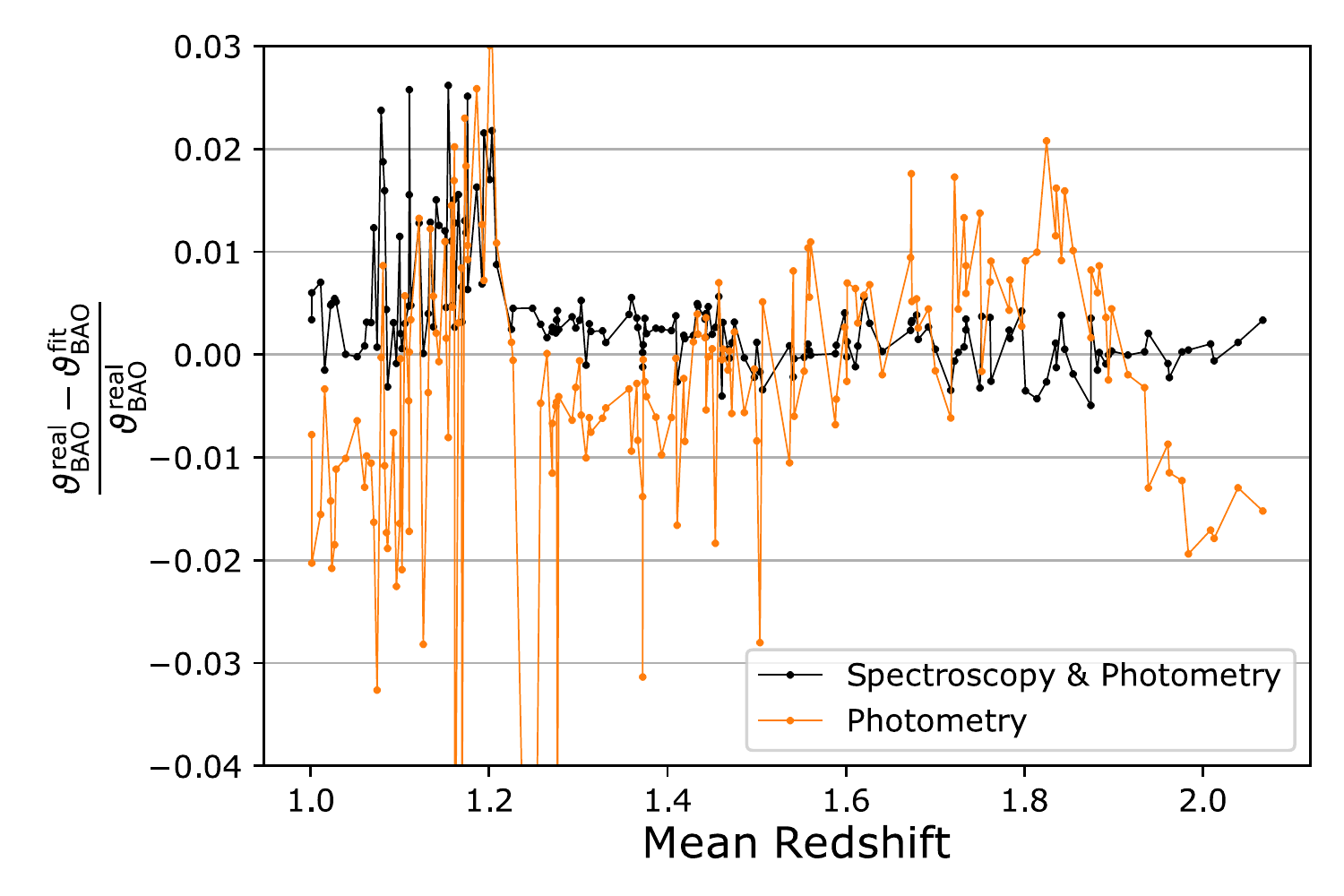}
                 \includegraphics[width = 0.5\textwidth]{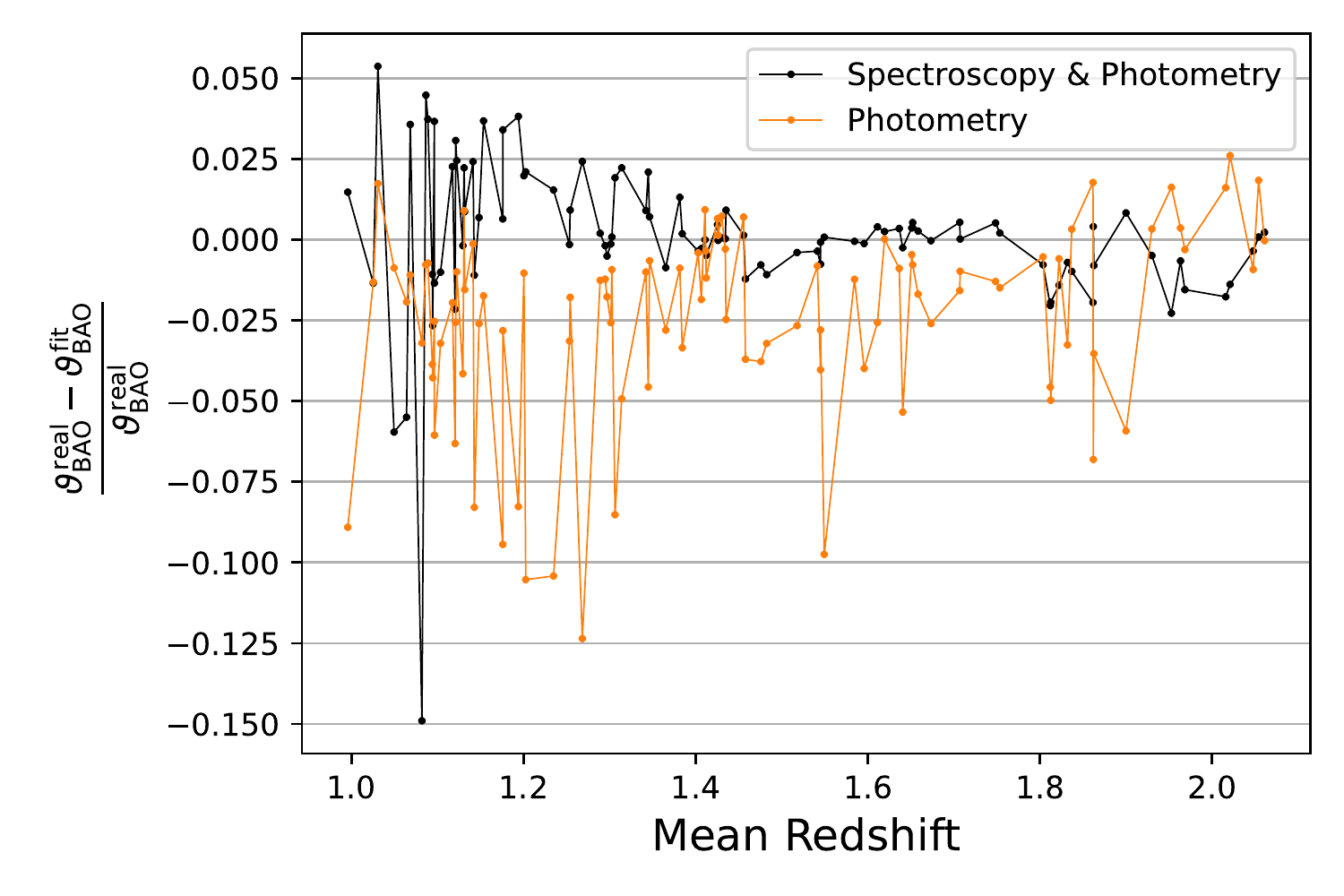}
                \caption{BAO relative errors as a function of redshift for the samples shown in Fig. \ref{fig:comparison-pz}. The results from the combination of stacked spectroscopy and photometry and stacked photometry are respectively plotted in black and orange. The BAO position could not be fitted in the spectroscopy-only analysis so the error in this case is not shown.}
                \label{fig:comparison-error}
            \end{figure}

            In Sect. \ref{introduction} we stated that the combination of stacked spectroscopy and photometry is needed in order to break the colour-redshift degeneracy and recover detailed redshift distributions. Here we justify this claim by comparing the results of different analyses that use the combination of stacked spectroscopy and photometry, stacked spectroscopy alone and stacked photometry.
            
            We analysed a subset of the non-attenuated catalogue and one of the real attenuation catalogue without measurement noise. The colour groups for the analyses were selected with the same criterion used for the analyses presented in the paper ($z_{\rm mean} > 1$ and $\sigma_z < 0.2$). In Figs. \ref{fig:comparison-pz} and \ref{fig:comparison-error} we present the results of the analysis that used the ElasticNet regularisation, which was the best performing linear regression method, with the best fitting parameters labelled as $\alpha_{\rm no \, noise}$ and $\beta_{\rm no \, noise}$ in Sect. \ref{parameter_selection}. Figure \ref{fig:comparison-pz} panels show two example redshift distribution fits, in the top panel for the non-attenuated catalogue and in the bottom for the real attenuation catalogue. From the plots it is clear that the analyses with stacked spectroscopy alone (green line) are not able to localise the peak of the redshift distribution. On the other hand, stacked photometry (orange line) is able to roughly locate the redshift distribution, but does not fit its substructure and presents spurious peaks.
            Finally, the combination of stacked spectroscopy and photometry (black line) breaks the colour-redshift degeneracies and recovers the redshift distribution with a significant improvement in accuracy that can be seen by eye.
            
            Figure \ref{fig:comparison-error} shows the BAO relative error derived for all of the colour groups with the combination of stacked spectroscopy and photometry, and for stacked photometry alone. We were unable to fit the BAO angular position for the analysis with stacked spectroscopy alone due to the disperse distributions that were recovered and so it is not shown on the plots. 
            The mean BAO error of the non-attenuated catalogue analyses is $0.0044$ for the combination of stacked spectroscopy and photometry and $0.010$ for the analyses with stacked photometry alone; for the real attenuation catalogue they are respectively $0.014$ and $0.028$. Thus, the addition of spectroscopy in the analysis reduces the error by a factor of 2.
            
            These results justify the choice of using the combination of stacked spectroscopy and photometry. Photometry is indeed needed in order to locate the redshift distribution, but the addition of spectroscopic information helps to break the degeneracies in colour-redshift space and significantly improves the constraints.
            \section{Colour division in SOMs} \label{colours-in-som}
            
            \begin{figure*}
                \centering
                \includegraphics[width = 1\textwidth]{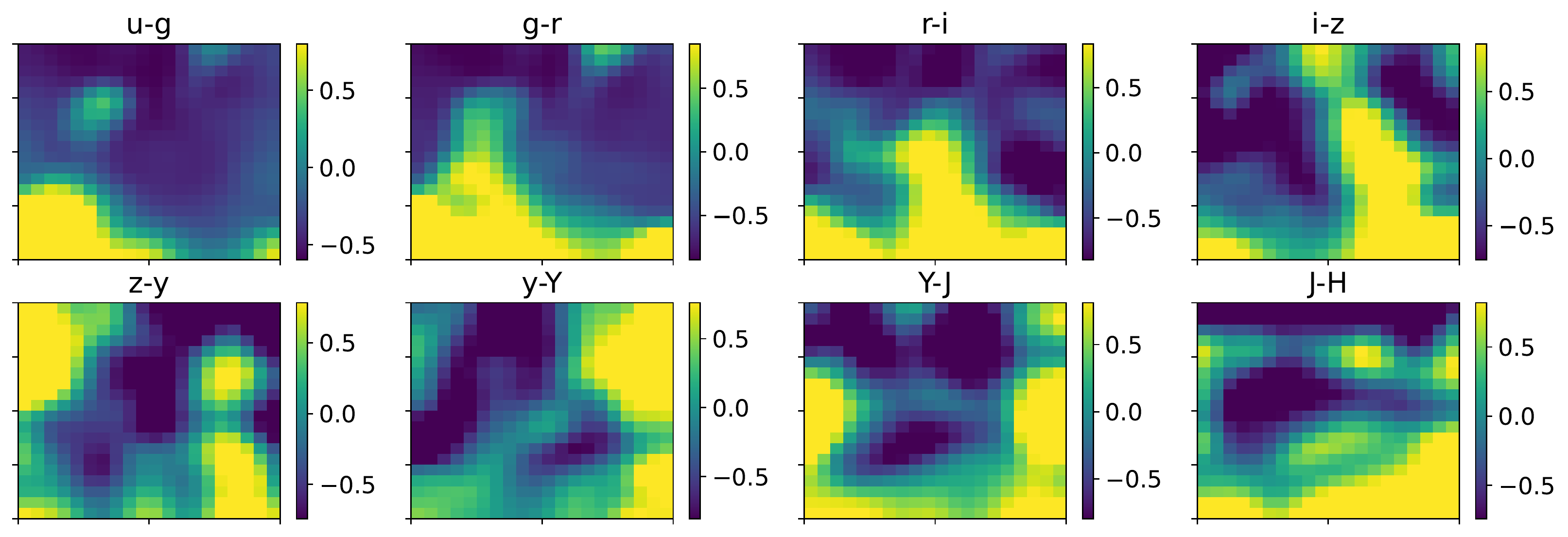}
                \caption{SOM shown on the left panel of Fig. \ref{fig:som}. The colour scale indicate the mean colour of the SOM cell, each panel represents one of the eight colours used to build the SOM.}
                \label{fig:colors-SOM}
            \end{figure*}
            
            \begin{figure*}
                \centering
                \includegraphics[width = 1\textwidth]{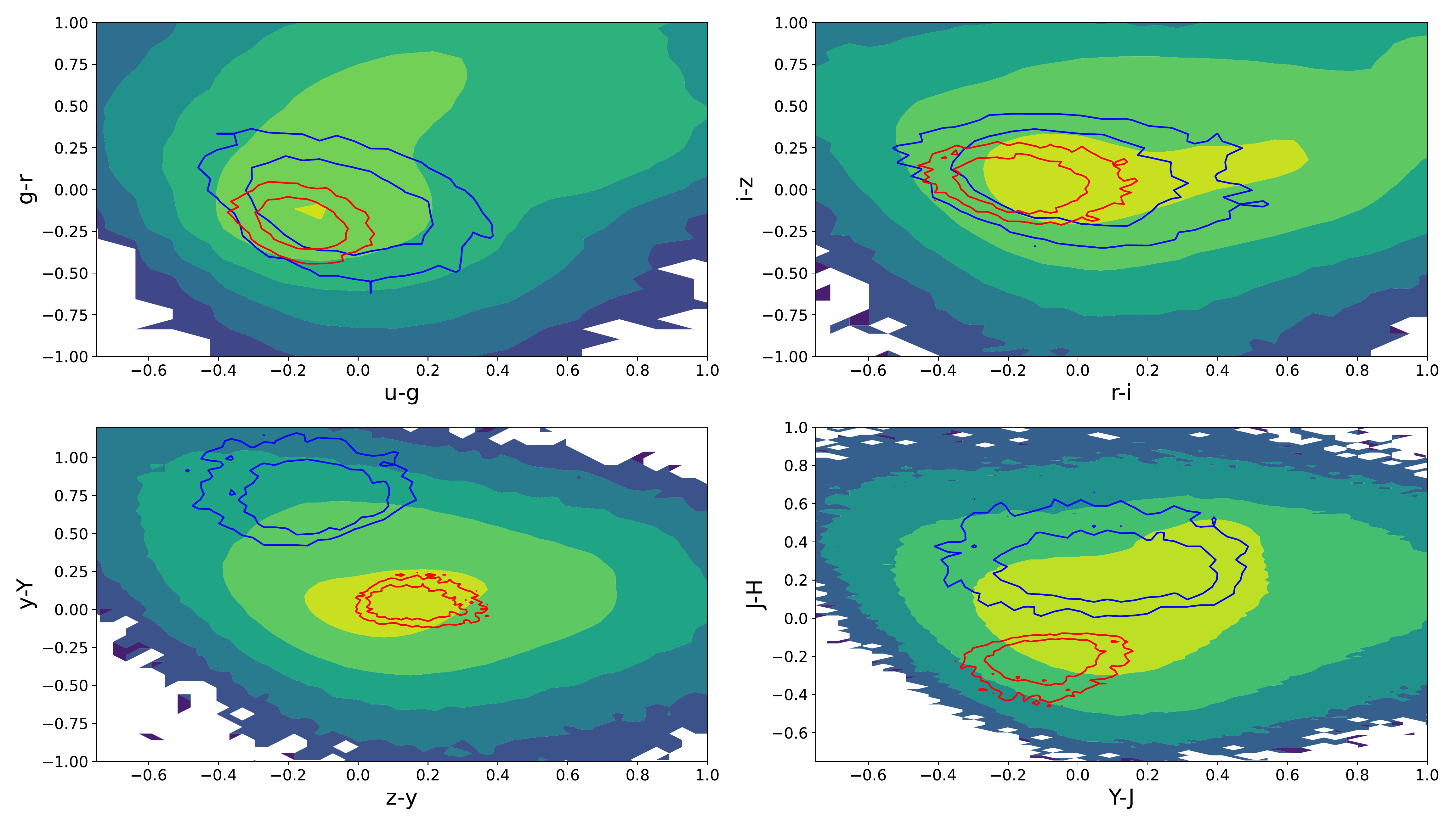}
                \caption{Contour plots of the colour distributions. The filled lines show the distribution of all galaxies in the catalogue, while the red and blue ones represent the colour distributions of two cells of the SOM. The inner and outer contour lines contain 90\% and 68\%  of the samples. The figure shows how the SOM groups galaxies with similar colours in the same cells and how different cells have different colour distributions.} 
                \label{fig:contour-colors}
            \end{figure*}
            
            In this work we used a very coarse SOM, as we need highly populated colour groups in order to average out the noise in the stacked spectrum (see Sect. \ref{sec:SOM} and Fig. \ref{fig:stack}). Figure \ref{fig:colors-SOM} shows how the colours are mapped to the SOM cells, and \ref{fig:contour-colors} shows the colour distributions of two  SOM cells. In Fig. \ref{fig:contour-colors} the colour distribution of two colour groups (blue and red contours) are over-plotted on the distribution of the whole catalogue. The distributions are compact in comparison to the catalogue one, showing that the SOM groups galaxies with similar colours. In addition, the two colour distributions are separated from one another in the bottom panels of Fig. \ref{fig:contour-colors}. This gives a first visual proof that even with a limited number of cells the SOM is able to divide the galaxies into distinct colour groups.
            
            \end{appendix}
            
            \end{document}